\documentclass[prb,preprint,onecolumn]{revtex4-1} 

\usepackage{amsmath}  
\usepackage{amsfonts} 
\usepackage{graphicx} 

\usepackage{tabularx} 
\usepackage{setspace}
\usepackage{subfig}
\usepackage{amssymb}

\begin{document}

\title{Haidinger-Michelson rings in white light}

\author{Youssef \textsc{El~Azhari}}
\email{youssef.elazhari@men.gov.ma}        

\affiliation{Centre national d'innovation p\'edagogique et d’exp\'erimentation (CNIPE)~-- Rabat (Morocco)}
\affiliation{Centre r\'egional des m\'etiers d’\'education et de formation (CRMEF)~-- Marrakesh (Morocco)}

\author{Sa\"\i d \textsc{Tagmouti}}
\email{said.tagmouti@men.gov.ma}
\affiliation{Centre r\'egional des m\'etiers d’\'education et de formation (CRMEF)~-- Casablanca (Morocco)}

\date{\today}

\begin{abstract}
We use coherence theory to explain why it is necessary to modify the conventional setup of a Michelson interferometer to obtain Haidinger rings with an extended source of white light. The modification consists of introducing a glass slide into one of the two arms of the interferometer. This insertion circumvents the drastic restriction imposed by the low temporal coherence of white light, which prevents the observation of interference rings with the conventional setup. In order to understand this restriction, we  developed and implemented a criterion for observing interference fringes. The modified setup also makes it possible to perform a spectral interferometry experiment to analyze the output of the interferometer and determine the refractive index of the glass slide over the whole visible spectrum. The fit of measured data using Sellmeier's law gives the extrapolated value of the refractive index to the IR $n_\mathrm{IR}=1.5307\pm0.0003$ and the value of the characteristic wavelength $\lambda_0=(164.5\pm0.4)~\mathrm{nm}$ of the oscillator responsible for the dispersion.
\end{abstract}

\maketitle 

\section{Introduction}
\begin{figure*}
\setlength\tabcolsep{2pt}%
\begin{tabularx}{\textwidth}{lcccc}
   &\begin{minipage}{0.23\textwidth}
    \textbf{\small Interfrometer lighting conditions}\vspace{0.2em}\null
    \end{minipage}
   &\begin{minipage}{0.23\textwidth}
    \textbf{\small Mirror arrangement}
    \end{minipage}
   &\begin{minipage}{0.23\textwidth}
    \textbf{\small Location of the observation screen}
    \end{minipage}
   &\begin{minipage}{0.23\textwidth}
    \textbf{\small Interference pattern}
    \end{minipage}
    \\ 
    \hline
    \begin{minipage}{0.04\textwidth}
    (a)
    \end{minipage}
   &\begin{minipage}{0.23\textwidth}
    \begin{spacing}{0.8}{\footnotesize Divergent beam from a point source.\\(Green HeNe Laser)}\end{spacing}
    \end{minipage}
   &\begin{minipage}{0.23\textwidth}
    \begin{spacing}{0.6}{\footnotesize Perpendicular}\end{spacing}
    \end{minipage}
   &\begin{minipage}{0.23\textwidth}
    \begin{spacing}{0.8}{\footnotesize\null Anywhere at the interferometer exit: the fringes are not localized.}\end{spacing}
    \end{minipage}
   &\begin{minipage}{0.23\textwidth}
    \null\vspace{0.1em}
    \includegraphics[width=0.6\linewidth,keepaspectratio]{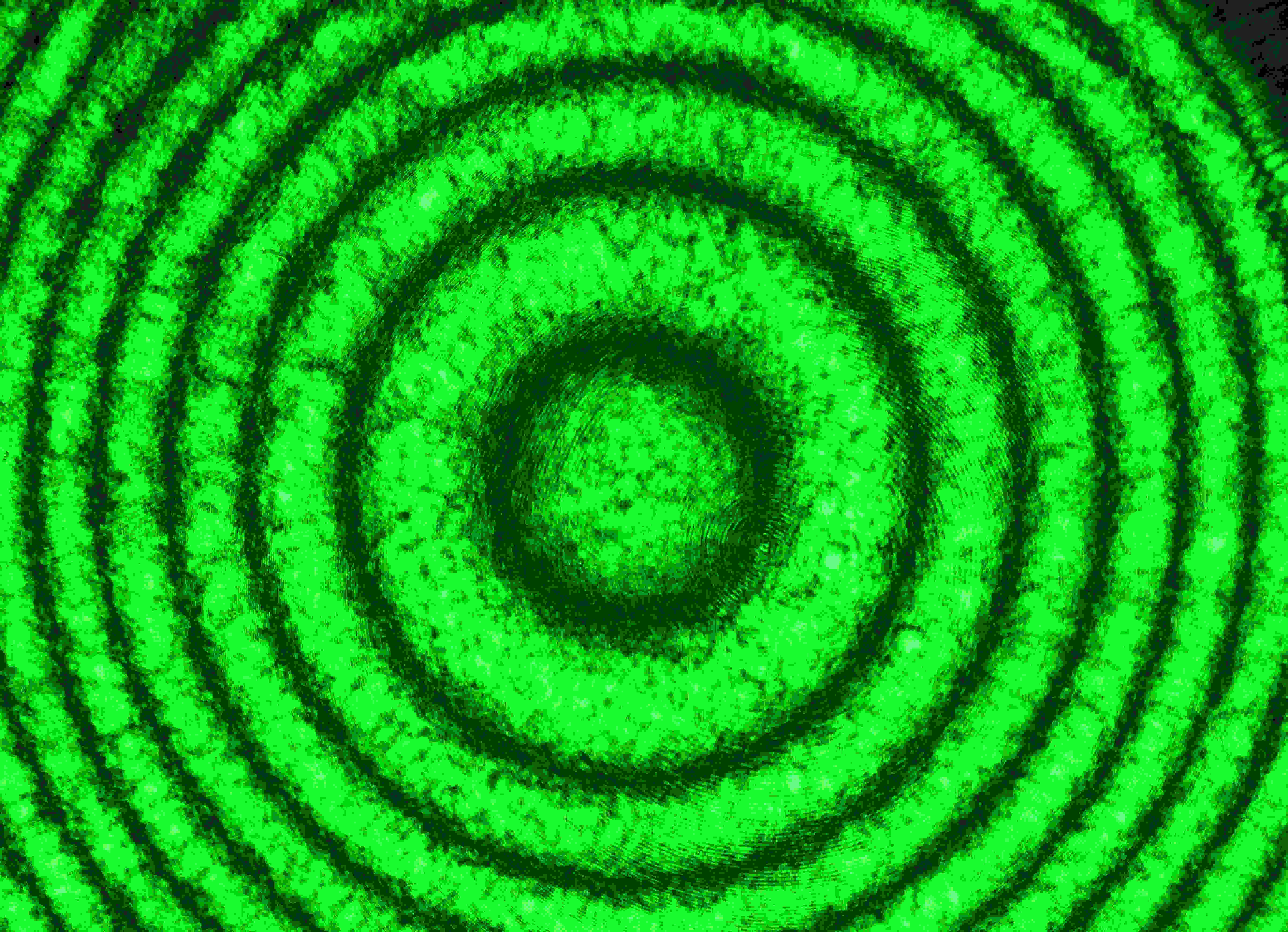}
    \vspace{0.1em}\null
    \end{minipage}
    \\
    \hline
    \begin{minipage}{0.04\textwidth}
    (b)
    \end{minipage}
   &\begin{minipage}{0.23\textwidth}
     \begin{spacing}{0.8}
     {\footnotesize Divergent beam from a point source.\\(Green HeNe Laser)}\end{spacing}
     \end{minipage}
   &\begin{minipage}{0.23\textwidth}
    \begin{spacing}{0.6}{\footnotesize Angled}\end{spacing}
    \end{minipage}
   &\begin{minipage}{0.23\textwidth}
    \begin{spacing}{0.8}
    {\footnotesize\null Anywhere at the interferometer exit: the fringes are not localized.}\end{spacing}
    \end{minipage}
   &\begin{minipage}{0.23\textwidth}
    \includegraphics[width=0.6\linewidth,keepaspectratio]{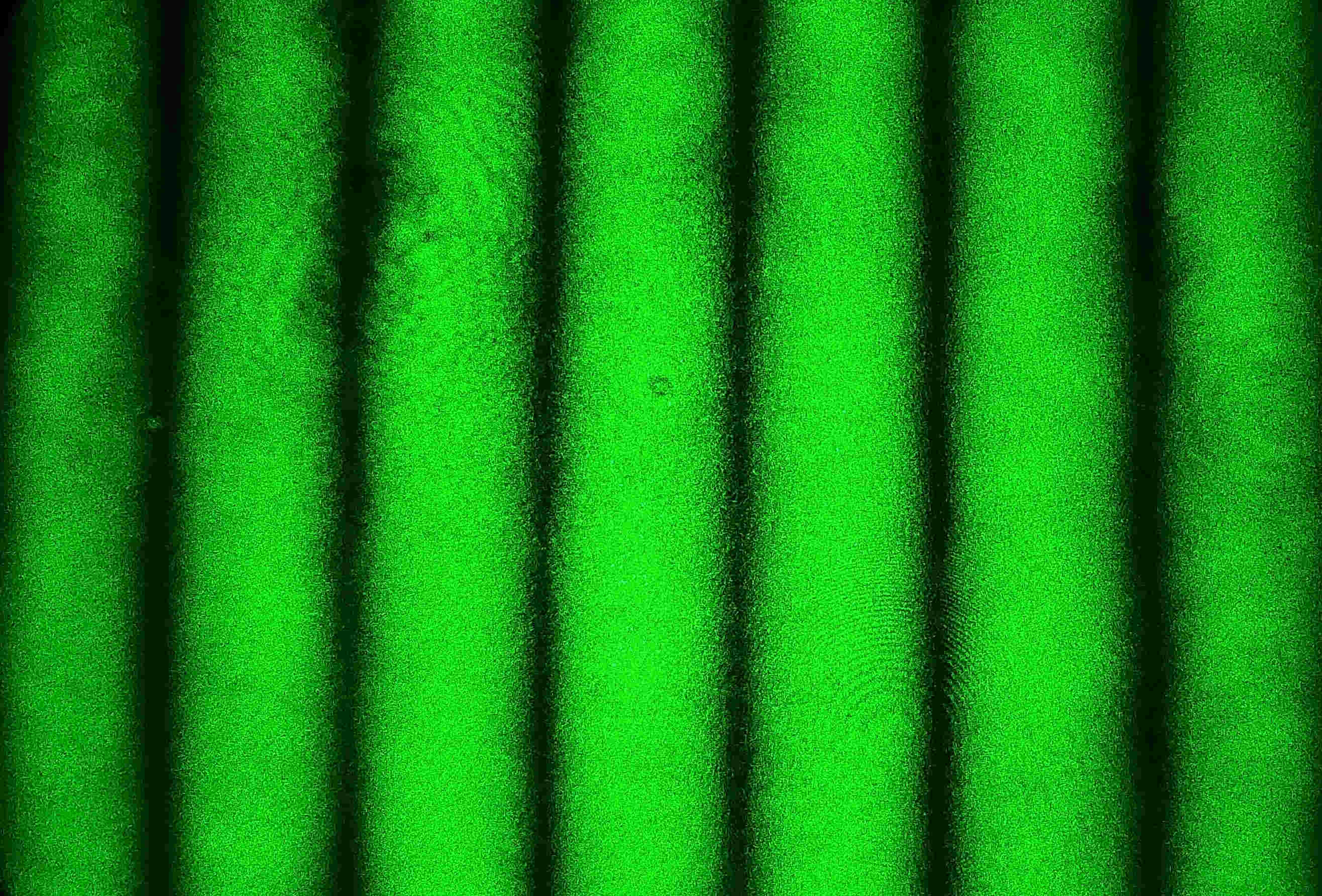}
    \end{minipage}
    \\
    \hline
    \begin{minipage}{0.04\textwidth}
    (c) 
    \end{minipage}
   &\begin{minipage}{0.23\textwidth}
     \begin{spacing}{0.8}
     {\footnotesize Parallel beam from a point source.\\(Red HeNe Laser)}
     \end{spacing}
     \end{minipage}
   &\begin{minipage}{0.23\textwidth}
    \begin{spacing}{0.6}
    {\footnotesize Angled}\end{spacing}
    \end{minipage}
   &\begin{minipage}{0.23\textwidth}
    \begin{spacing}{0.8}
    {\footnotesize\null Anywhere at the interferometer exit: the fringes are not localized.}
    \end{spacing}
    \end{minipage}
   &\begin{minipage}{0.23\textwidth}
    \includegraphics[width=0.6\linewidth,keepaspectratio]{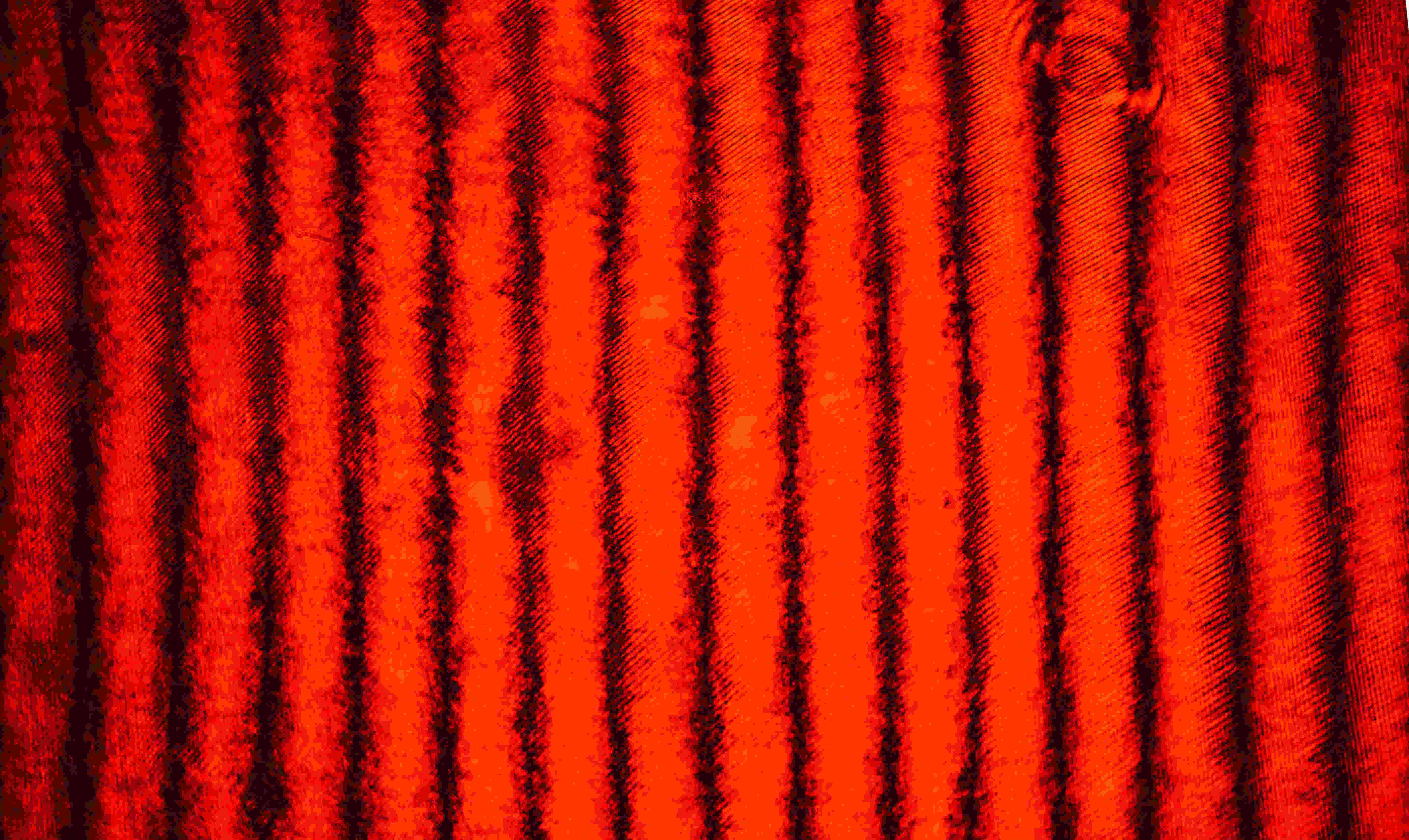}
    \end{minipage}
    \\
    \hline
    \begin{minipage}{0.04\textwidth}
    (d)
    \end{minipage}
   &\begin{minipage}{0.23\textwidth}
     \begin{spacing}{0.8}
     {\footnotesize\null Almost parallel beam from an extended sodium spectral source.}
     \end{spacing}
     \end{minipage}
   &\begin{minipage}{0.23\textwidth}
    \begin{spacing}{0.8}
    {\footnotesize\null Angled}
    \end{spacing}
    \end{minipage}
   &\begin{minipage}{0.23\textwidth}
    \begin{spacing}{0.8}
    {\footnotesize\null Conjugated to the air wedge by a converging lens placed at the exit of the interferometer.}
    \end{spacing}
    \end{minipage}
   &\begin{minipage}{0.23\textwidth}
    \includegraphics[width=0.6\linewidth,keepaspectratio]{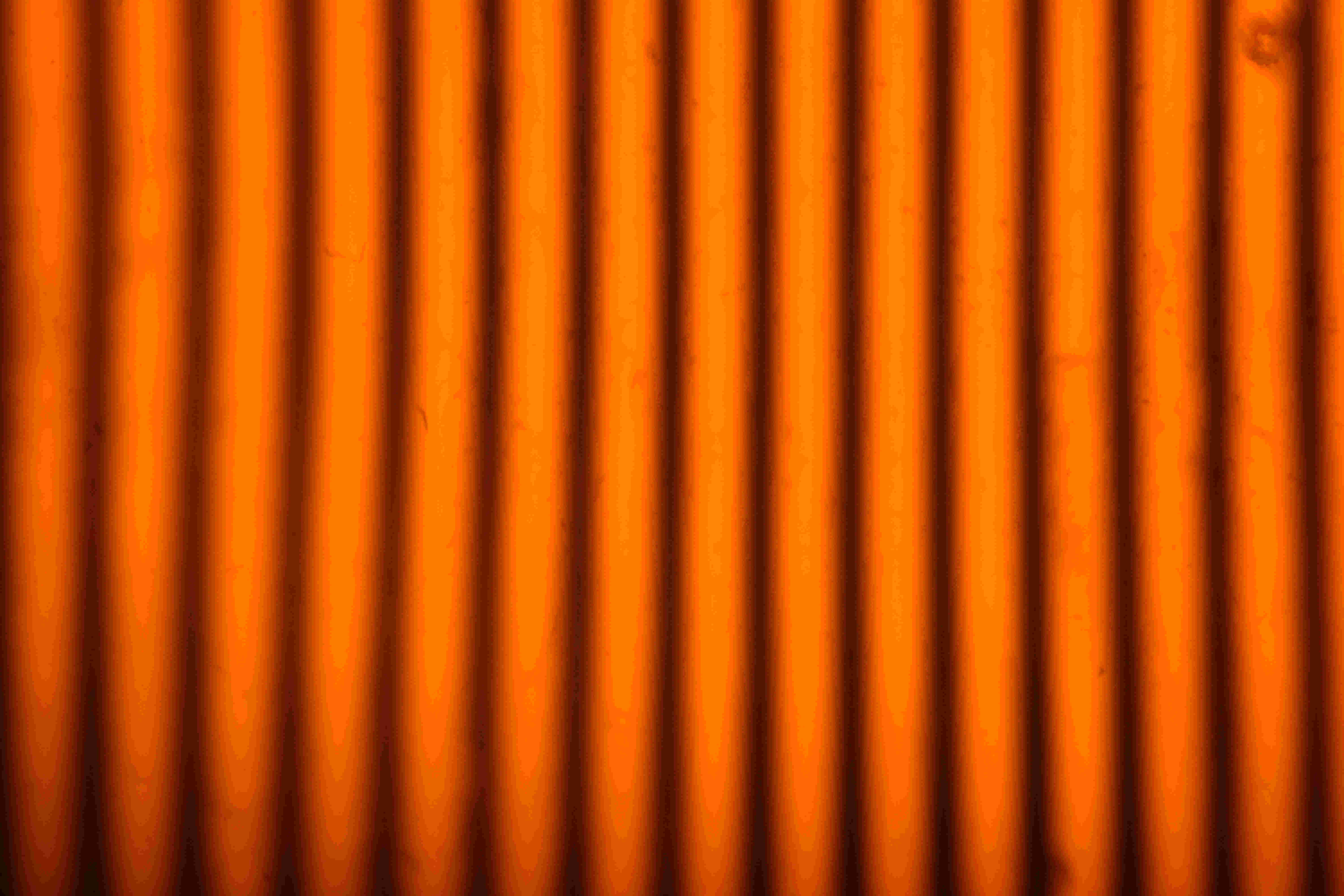}
    \end{minipage}
   \\ \hline
    \begin{minipage}{0.04\textwidth}
    (e)
    \end{minipage}
   &\begin{minipage}{0.23\textwidth}
    \begin{spacing}{0.8}{\footnotesize\null Almost parallel beam from an extended white source.}\end{spacing}
    \end{minipage}
   &\begin{minipage}{0.23\textwidth}
    \begin{spacing}{0.8}
    {\footnotesize\null Angled}
    \end{spacing}
    \end{minipage}
   &\begin{minipage}{0.23\textwidth}
    \begin{spacing}{0.8}
    {\footnotesize\null Conjugated to the air wedge by a converging lens placed at the exit of the interferometer.}
    \end{spacing}
    \end{minipage}
   &\begin{minipage}{0.23\textwidth}
    \includegraphics[width=0.6\linewidth,keepaspectratio]{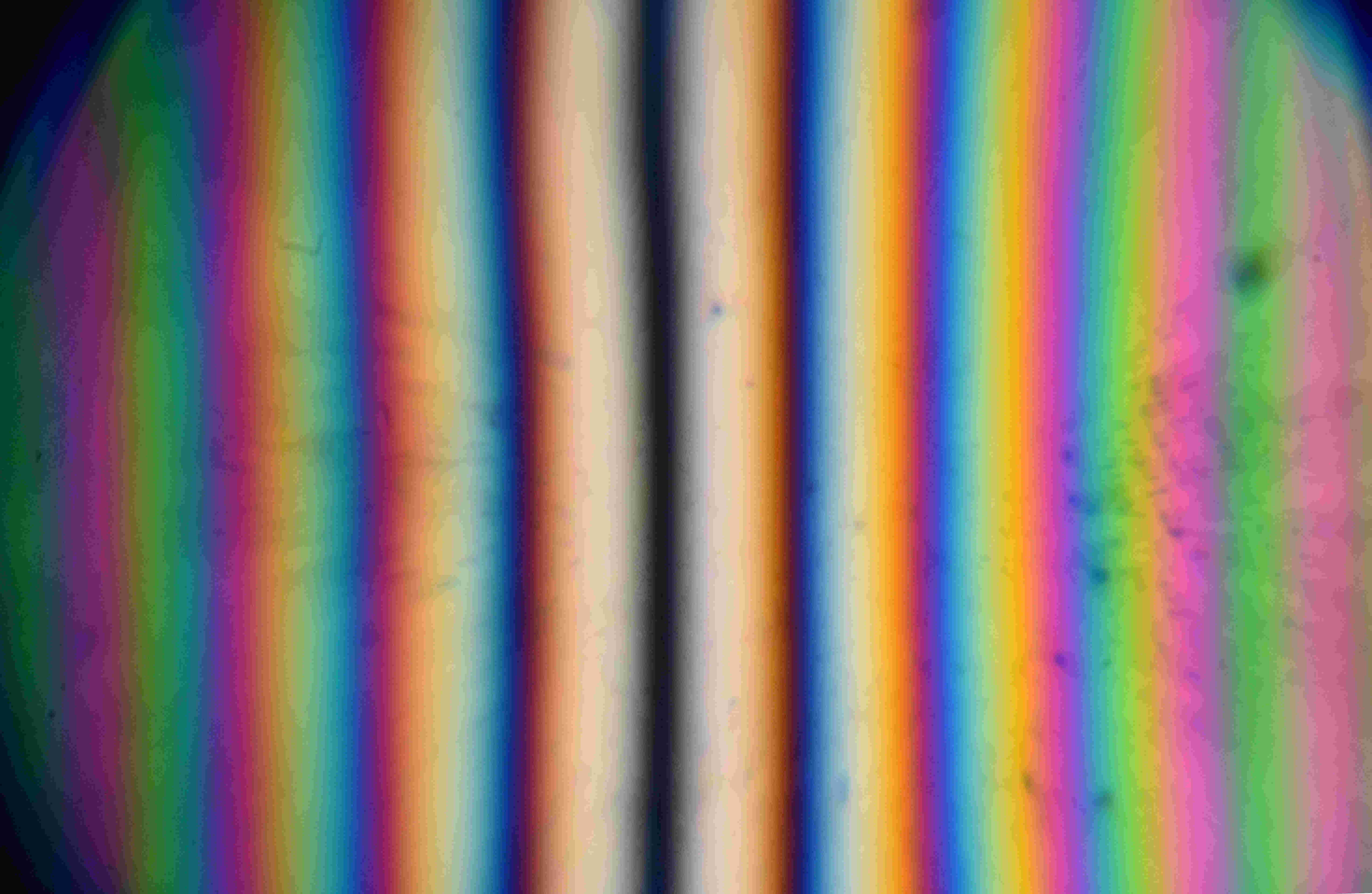}
    \end{minipage}
    \\ 
    \hline
    \begin{minipage}{0.04\textwidth}
    (f)
    \end{minipage}
   &\begin{minipage}{0.23\textwidth}
     \begin{spacing}{0.8}
     {\footnotesize\null Convergent beam from an extended sodium spectral source.}
     \end{spacing}
     \end{minipage}
   &\begin{minipage}{0.23\textwidth}
    \begin{spacing}{0.8}
    {\footnotesize\null Perpendicular}
    \end{spacing}
    \end{minipage}
   &\begin{minipage}{0.23\textwidth}
    \begin{spacing}{0.8}
    {\footnotesize\null Focal plane of a converging lens placed at the exit of the interferometer.}
    \end{spacing}
    \end{minipage}
   &\begin{minipage}{0.23\textwidth}
    \includegraphics[width=0.6\linewidth,keepaspectratio]{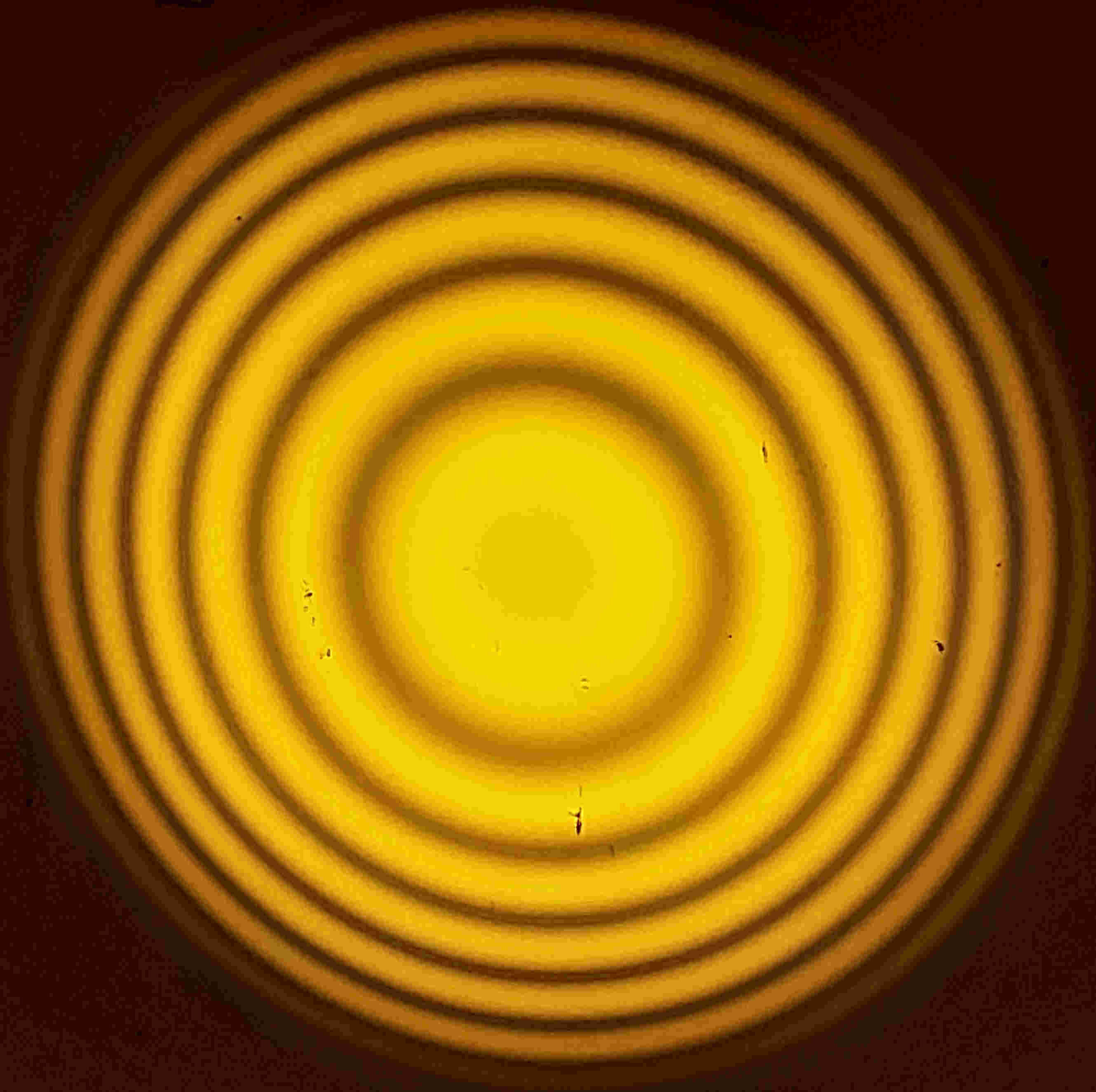}
    \end{minipage}
\end{tabularx}
\caption{Interference patterns obtained by using different setups based on a Michelson interferometer. 
         \label{fig:DifferentsFiguresInterferenceMichelson}}
\end{figure*}

The Michelson interferometer plays an important role in teaching experimental physics at the undergraduate level in universities throughout the world. This device makes it possible to carry out a whole range of optical interference experiments such as:
\begin{list}{--}{}
\item non-localized interference from a point source,\citep{Gasvik2002} such as a laser, placed at a finite distance;\citep{Lipson2011,Fang2013} in this case, interference fringes are either concentric rings (Fig.~\ref{fig:DifferentsFiguresInterferenceMichelson}(a)) or hyperbolae which can approximate equidistant straight line segments near their centers (Fig.~\ref{fig:DifferentsFiguresInterferenceMichelson}(b));
\item non-localized interference between two plane waves,\citep{ElAzhari2005} produced, for example from a laser source whose beam is enlarged by means of an afocal telescope, giving rectilinear fringes (Fig.~\ref{fig:DifferentsFiguresInterferenceMichelson}(c));
\item rectilinear localized fringes or fringes of equal thickness, also called Fizeau fringes,\citep{Jenkins2001,Grievenkamp2002} obtained either from an extended spectral source (Fig.~\ref{fig:DifferentsFiguresInterferenceMichelson}(d)) or a \emph{white light} extended source (Fig.~\ref{fig:DifferentsFiguresInterferenceMichelson}(e)) illuminating the Michelson interferometer with an almost parallel light beam;
\item circular localized fringes or fringes of equal inclination, also called Haidinger rings,\citep{Jenkins2001,Grievenkamp2002} obtained from an extended spectral source illuminating the Michelson interferometer with a converging light beam (Fig. \ref{fig:DifferentsFiguresInterferenceMichelson}(f)).
\end{list}

\par
\begin{figure*}[t]
  \begin{center}
  \hfil
  \subfloat[$e=1.0\;\mathrm{\mu m}$. \label{subfig:ContactOptUnMicroQuatre}]
           {\includegraphics[height=4cm]{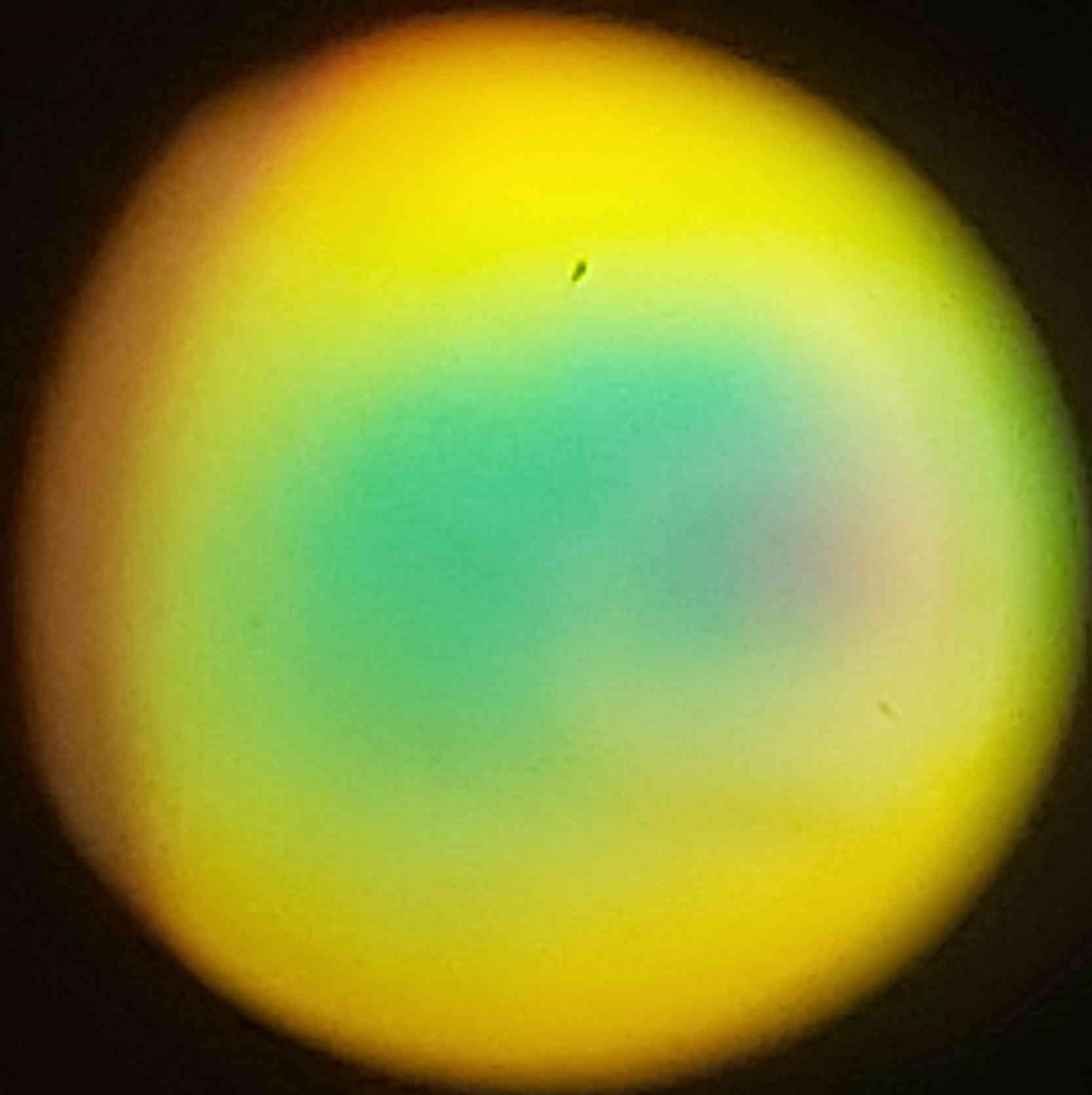}}
  \hspace{0.03\textwidth}
  \subfloat[$e=2.0\;\mathrm{\mu m}$. \label{subfig:ContactOptDeuxMicroZero}]
           {\includegraphics[height=4cm]{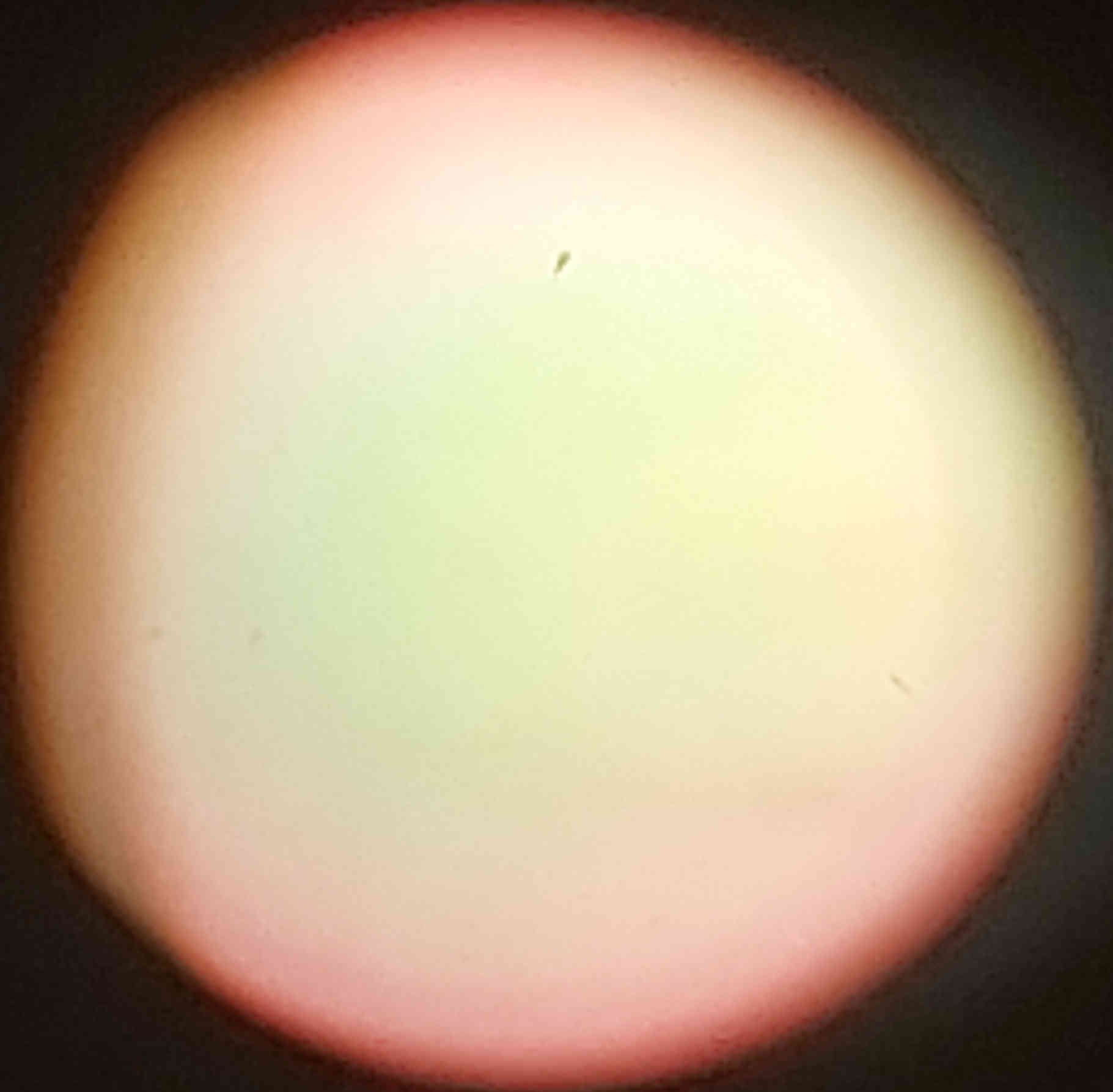}}
  \hspace{0.03\textwidth}
  \subfloat[$e=3.0\;\mathrm{\mu m}$. \label{subfig:ContactOptTroisMicroZero}]
           {\includegraphics[height=4cm]{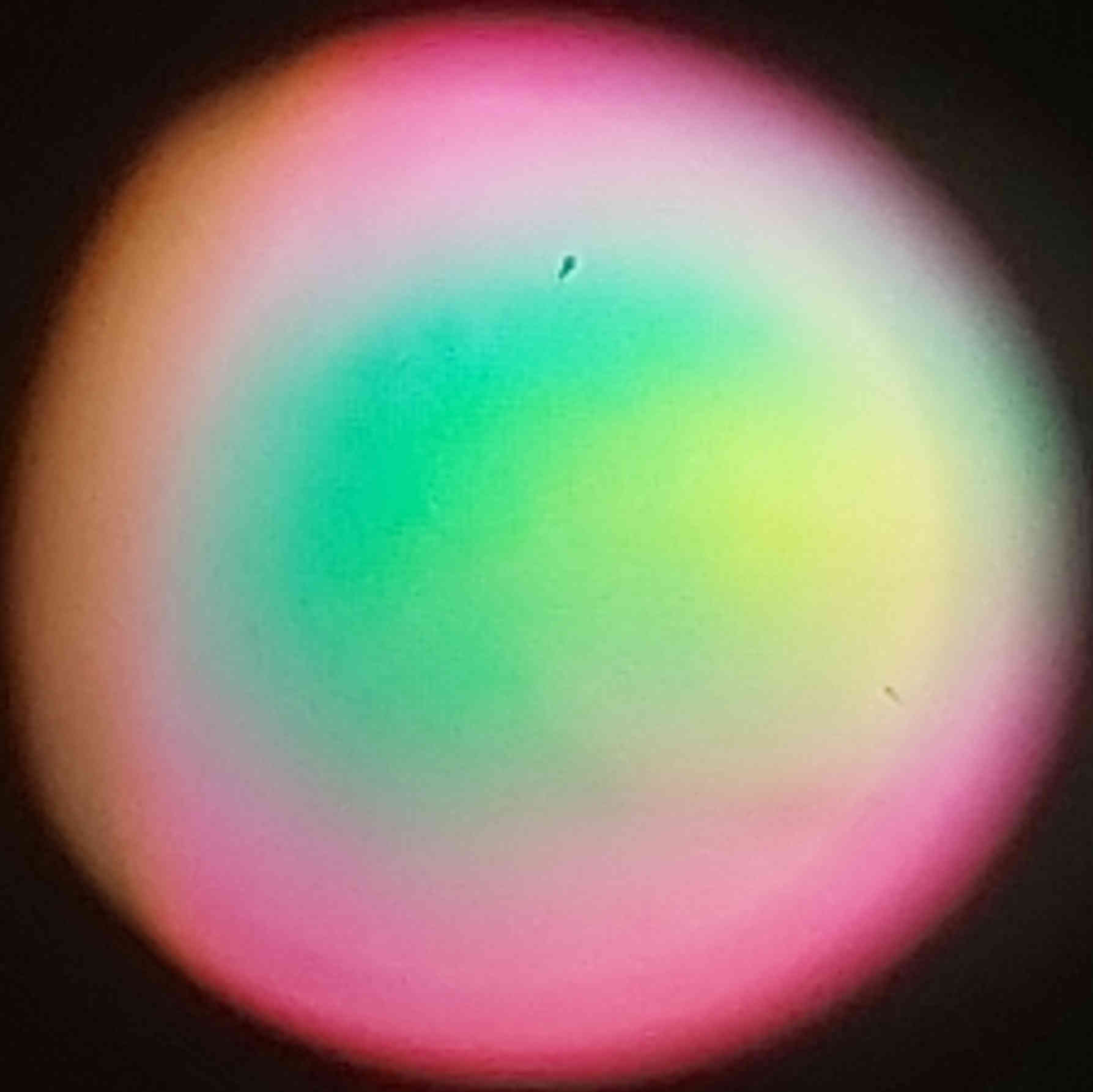}}
  \hfil
  \mbox{}
  \caption{Interference patterns obtained with a Michelson interferometer configured into a parallel-faces air blade for 
           three different thicknesses $e=|\ell_1-\ell_2|$ of the equivalent parallel-faces air blade; $\ell_1$ and $
           \ell_2$ are the lengths of the two arms of the interferometer. \label{fig:EtatEcranContactOptique}}
  \end{center}
\end{figure*}

 However, it is extremely difficult to obtain Haidinger rings in white light with the Michelson interferometers found in most teaching laboratories. Figure \ref{fig:EtatEcranContactOptique} shows the observation screen when trying to obtain such fringes using a conventional setup based on a Michelson interferometer. The photographs correspond to three different distances of the movable mirror from the point of optical contact. The two mirrors of a Michelson interferometer are said to be in optical contact when the lengths of the two arms of the interferomer are equal. As soon as one moves a mirror far enough from the point of optical contact to reveal any interference fringes, the observation screen becomes uniformly illuminated with the same color as the source.

\par To explain the experimental difficulty of obtaining Haidinger rings in white light with typical Michelson interferometers, we will adopt a practical criterion for observing fringes (section~\ref{sec:HaidingerClassique}). Then, we show (section~\ref{sec:HaidingerModifie}) how a very simple modification of the setup enables the observation of Haidinger-Michelson rings in white light. A protocol is also given for carrying out such an experiment.

\par The modification consists of introducing a glass slide with parallel faces into one of the two arms of the interferometer. The effect of such a glass slide on the interference fringes was first presented by Birchby\citep{Birchby1924} and discussed by Sethi\citep{Sethi1924} and Birchby.\citep{Birchby1924,Birchby1927} Later on, Young and O'Connor\citep{Young1970} gave a theoretical explanation of the phenomenon using the notion of an achromatic fringe while Zwick and Shephered\citep{Zwick1971} applied it in a wide-angle Michelson interferometer.

\par Measurement of the refractive index of optical media is of practical interest and such measurements have been performed using different methods.\citep{Honijk1973,ElAzhari2000} However these methods are difficult to implement and often call on sophisticated theories. Obtaining Haidinger-Michelson fringes opens the way to an extremely simple, precise and rapid method of measuring the refractive index of a transparent medium over the whole visible spectrum. This consists (section~\ref{sec:InterferometrieSpectrale}) of implementing a spectral interferometry experiment\citep{Froehly1973,Kovacs1998,Dorrer2000} and developing a method for calculating the index of refraction from the spectral fringes.

\par\medskip Pedagogically speaking, this article is mainly intended for teachers of undergraduate levels who would like to: i) explain clearly and simply why one cannot obtain Haidinger rings in white light using a Michelson interferometer and/or ii) quantitatively and very simply perform and analyze a spectral interferometry experiment with teaching laboratory equipment to extract the wavelength dependence of the refractive index of a glass slide throughout the whole visible spectral range.

\par For this, we assume that readers are familiar with: i) the general concept of two-wave light interference, ii) the properties of lenses and the principles of image formation and iii) the effect of temporal coherence on wave interference. Other necessary concepts for understanding this article can be found in many optics textbooks\citep{Hecht2002,Born1998,Sharma2006,Marchenko2007} and will gradually be introduced hereafter.

\section{conventional setup\label{sec:HaidingerClassique}}
The Michelson interferometer is often used to obtain Haidinger rings\citep{Hecht2002,Raman1939} or fringes of equal inclination, localized at infinity. The purpose of this section is to discuss how the conventional setup works and to explain why it is impossible to produce Haidinger rings using such a setup.

\subsection{Brief description of the conventional setup\label{ssec:BreveDescriptionMontageClassique}}
\begin{figure*}[ht]
  \begin{center}
  \includegraphics[scale=1]{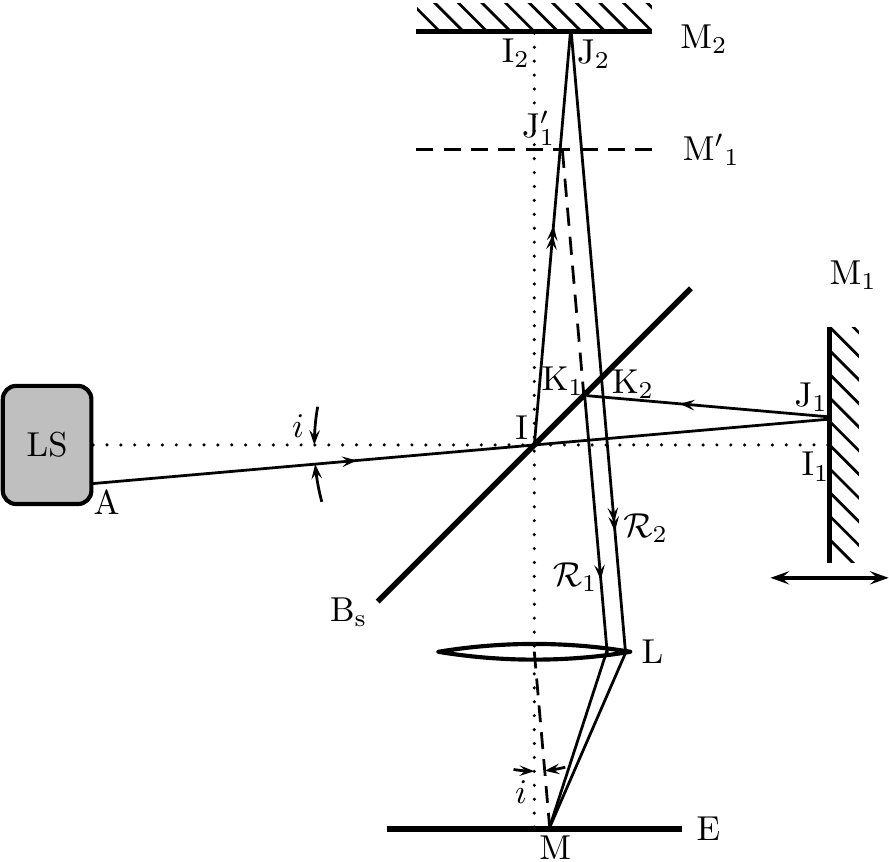}
  \caption{Conventional setup to obtain the Haidinger rings using a Michelson interferometer with two orthogonal mirrors $
           \mathrm{{M_1}}$ and $\mathrm{{M_2}}$. $\text{LS}$ is an extended light source; $\mathrm{B_s}$ represents the 
           beam splitter; $\mathrm{{M_1}}$ is the moving mirror; and $\mathrm{M_2}$ the fixed mirror. $\mathrm{L}$ is a 
           converging lens which forms the image of the interference pattern on the observation screen $\mathrm{E}$ placed 
           in the image focal plane of $\mathrm{L}$. The image $\mathrm{M}'_1$ is explained in 
           Fig.~\ref{fig:LameDairEquivalenteMichelson}.\label{fig:MontageClassiqueMichelsonLameDair}}
  \end{center}
\end{figure*}

The conventional setup uses an extended source of light to illuminate the Michelson interferometer with a convergent light beam. One can represent such an interferometer by two plane mirrors $\mathrm{M_1}$ and $\mathrm{M_2}$ and a $50$-$50$ beam splitter $\mathrm{B_s}$ placed at $45^\circ$ to the mean direction of propagation of the incident beam as shown in Fig.~\ref{fig:MontageClassiqueMichelsonLameDair}.

\par The mirrors $\mathrm{M_1}$ and $\mathrm{M_2}$ are arranged perpendicular to each other. The mirror $\mathrm{M_1}$ is mounted on a translation stage.

\par In the case of an extended source, the interfering rays ${\cal R}_1$ and ${\cal R}_2$ come from the same incident ray $\mathrm{AI}$ reflecting from mirrors $\mathrm{M_1}$ and $\mathrm{M_2}$. Figure \ref{fig:MontageClassiqueMichelsonLameDair} indicates the paths followed by these rays inside the interferometer. They are parallel after leaving the interferometer and overlap at infinity. This is why the interference pattern is located at infinity. In practice, fringes can be observed on a screen placed far enough from the interferometer or in the focal plane of a converging lens (see Fig. \ref{fig:MontageClassiqueMichelsonLameDair}).

\subsection{Interference pattern\label{ssec:ExpressionIllumination}}
\begin{figure*}[ht]
  \begin{center}
  \includegraphics[scale=1]{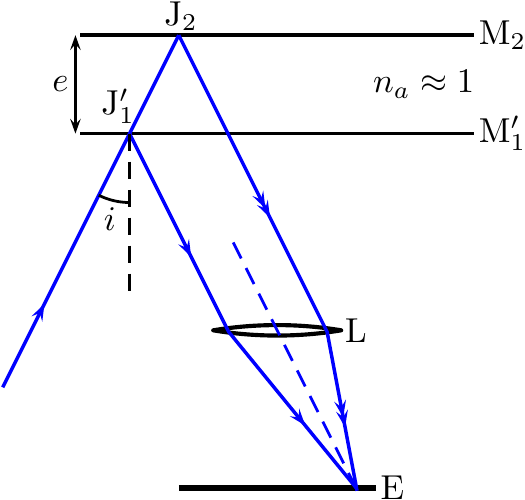}
  \caption{Scheme of the parallel-faces air blade equivalent to the conventional setup, making it possible to obtain the 
           Haidinger rings using a Michelson interferometer.\label{fig:LameDairEquivalenteMichelson}}
  \end{center}
\end{figure*}

Figure \ref{fig:MontageClassiqueMichelsonLameDair} also shows the image $\mathrm{M'_1}$ of the mirror $\mathrm{M_1}$ produced by the beam splitter $\mathrm{B_s}$. Because the optical paths between the points $\mathrm{A}$ and $\mathrm{M}$ are the same along the rays $\mathrm{AIJ_1'K_1M}$ and $\mathrm{AIJ_1K_1M}$, then such an arrangement of the Michelson interferometer is equivalent to a parallel-faces air blade (see Fig. \ref{fig:LameDairEquivalenteMichelson}) of index $n_a\approx1$ and thickness $e=|\ell_2-\ell_1|$, where $\ell_1$ and $\ell_2$ denote the lengths of the two arms of the interferometer.

\par Usually, the interferometer is illuminated with a convergent beam on the mirrors. The monochromatic irradiance of wavelength $\lambda$, obtained on a screen $\mathrm{E}$ placed in the plane of fringe localization, is given by:
\begin{align}
\label{eq:ExpressionEclairementMonochromatiqueMichelsonClassique}
I(\delta)=2\,I_0(\lambda)\,\left[{1-\cos\left({2\,\pi\,\frac{\delta}{\lambda}}\right)}\right],
\end{align}
where $\delta$ is the optical path difference between the two waves that are interfering, given by:
\begin{align}
\label{eq:DifferenceDeMarcheGeometriqueMichelsonClassiqueLameDair}
\delta=2\,e\cos(i),
\end{align}
and $i$ the angle of incidence on the mirrors. The minus sign that appears in Eq.~\eqref{eq:ExpressionEclairementMonochromatiqueMichelsonClassique} is due to the extra phase difference of $\pi$ introduced by the reflections from the beam splitter.\citep{Born1998} 

\par In the current case, the fringes correspond to $i=\text{constant}$. They are rings located at infinity. In practice, they can be observed at great distance from the interferometer or on a screen placed in the focal plane of a converging lens $\mathrm{L}$ located at the exit of the interferometer.

\par In the context of the paraxial approximation ($i\ll1$), the angular radius $i_m(\lambda)$ of the $m\;\text{th}$ dark ring is given by:
\begin{align}
i_m(\lambda)=\sqrt{\frac{\lambda}{e}\,(m+\epsilon)},
\end{align}
where
\begin{align}
\epsilon=p_0-\mathrm{E}(p_0).
\end{align}
$\mathrm{E}(p_0)$ is the integer part of the interference order $p_0=2\,e/\lambda$ at the center of the interference pattern ($i=0$).

\par With white light illumination, the situation is a little more complex. Indeed, not being mutually coherent, the different radiations that compose the white light do not interfere with each other. Thus, the intensity obtained at each point of the fringe localization plane is the result of the superposition of the different interferential systems:
\begin{align}
\label{eq:ExpressionEclairementLumiereBlancheMichelsonClassique}
I(\delta)=2\,\int_{\lambda_\text{min}}^{\lambda_\text{max}} I_\lambda^0(\lambda)\,\left[{1-\cos\left({2\,\pi\,\frac{\delta}{\lambda}}\right)}\right] \mathrm{d}\lambda.
\end{align}
The analysis of this expression is not so easy. It requires taking into account the temporal coherence of the light and the properties of the detector used (monochrome or trichromatic type, etc.). For this and in all cases, a criterion of observation of the fringes in polychromatic light becomes necessary. Before developing such a criterion, we introduce some notions about temporal coherence.

\subsection{Temporal coherence\label{ssec:NotionSuccincteCoherenceTemporelle}}
In a simple model of wave trains,\citep{Fowles1989} two overlapping secondary wave trains can interfere if they come from the same primary wave train. This requires that the optical path difference $\delta$ must be less than the average length of the wave trains $L_\tau=c\,\tau$, where $c$ is the speed of light and $\tau$ the average duration of wave trains.

\par Moreover, considering the properties of the Fourier transform, the average duration $\tau$ of the wave trains can be connected to the frequency width $\Delta\nu$ of the radiation by the characteristic relation:\citep{Born1998}
\begin{align}
\tau\;\Delta\nu\geqslant\frac{1}{4\,\pi}.
\end{align}
At the limit, we can take as expression of the temporal coherence length:
\begin{align}
L_\tau=\frac{\lambda^2}{4\,\pi\,\Delta\lambda}.
\end{align}
A practical condition for observing interference fringes could then be written:
\begin{align}
\label{eq:ExpressionGalContrainteDdmCoherenceTemporelle}
|\delta|\leqslant\eta\,L_\tau,
\end{align}
$\eta$ being a numerical factor of the order of a few units to a few tens, according to the desired accuracy. In the following, we will typically take $\eta=40$. This value seems to best correspond to our experiments. Table \ref{tab:LogueurTrainOndesDifferentesSources} gives the values of the average length of the wave trains $L_\tau$ for some visible radiation sources of average wavelength $\lambda$ and spectral width $\Delta\lambda$.

\begin{table*}[ht]
  \begin{center}
  \begin{ruledtabular}
    \begin{tabular}{rccc}
                        &$\lambda\mathrm{(nm)}$&$\Delta\lambda\mathrm{(nm)}$&$L_\tau\mathrm{(\mu m)}$\\
    \hline
    White light         &550                   &300                         &0.1\\
    Interference filter &550                   &10                          &2.4\\
    Spectral line       &550                   &1                           &24.1
    \end{tabular}
    \caption{Typical length of the wave trains $L_\tau$ for some visible radiation sources of average wavelength $\lambda$ 
             and spectral width $\Delta\lambda$.\label{tab:LogueurTrainOndesDifferentesSources}}
  \end{ruledtabular}
  \end{center}
\end{table*}

\subsection{Fringe visibility criterion in white light}
In order to be able to observe the equal inclination fringes or Haidinger rings with the conventional setup of Fig.~\ref{fig:MontageClassiqueMichelsonLameDair}, two conditions must be fulfilled simultaneously.

\par
\begin{figure*}[ht]
  \begin{center}
    \includegraphics[scale=1]{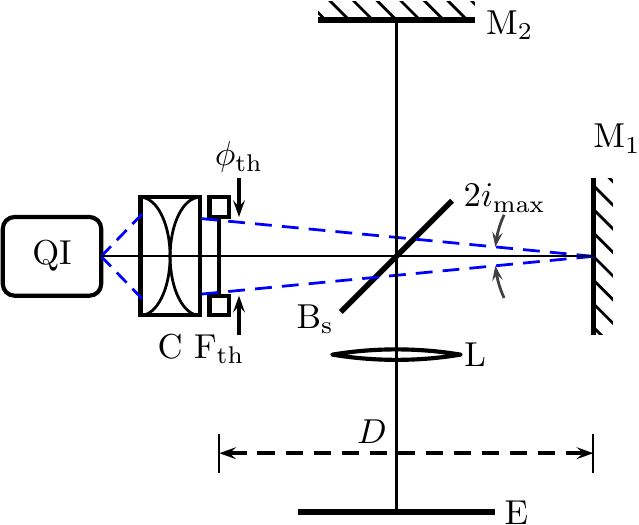}
    \caption{Michelson interferometer, in orthogonal configuration, illuminated by a converging beam of a quartz-iodine 
             lamp ($\mathrm{QI}$). $\mathrm{C}$ is a condenser or a converging lens; $\mathrm{F_{th}}$ is an optional
             thermal filter of diameter $\phi_\text{th}$.\label{fig:MichelsonCondenseurFiltreAntitehrmique}}
  \end{center}
\end{figure*}

 First, the extent of the field of observation is necessarily limited and must satisfy $i\leqslant i_\text{max}$, where $i_\text{max}$ is the maximum value of the angle of incidence allowed by the experimental setup. In our case, the field of observation is limited by the aperture stop constituted by the support of the thermal filter $\mathrm{F_{th}}$ (see Fig. \ref{fig:MichelsonCondenseurFiltreAntitehrmique}) and $i_\text{max}$ is given by:
\begin{align}
\label{eq:AngleOuvertureMaximumMichelson}
i_\text{max}=\arctan\left({\frac{\phi_\text{th}}{2\,D}}\right),
\end{align}
where $\phi_\text{th}$ is the opening diameter of the thermal filter placed at the entrance of the interferometer (see Fig. \ref{fig:MichelsonCondenseurFiltreAntitehrmique}) at a distance $D$ from the mirror $\mathrm{M}_1$. For a Michelson interferometer with $\phi_\text{th}=4\;{\rm cm}$ and $D=20\;{\rm cm}$, Eq.~\eqref{eq:AngleOuvertureMaximumMichelson} yields $i_\text{max}=0.1$.

\par Second, the optical path difference $\delta$ must satisfy the ``coherence condition" of Eq.~\eqref{eq:ExpressionGalContrainteDdmCoherenceTemporelle}. This condition imposes a maximum limit for the optical path difference and consequently for the thickness of the equivalent air-blade. Considering Eq.~\eqref{eq:DifferenceDeMarcheGeometriqueMichelsonClassiqueLameDair} and $\cos(i)\approx1$, Eq.~\eqref{eq:ExpressionGalContrainteDdmCoherenceTemporelle} yields:
\begin{align}
\label{eq:EpaisseurMaxMontageClassiqueHaidinger}
e_\text{max}=\frac{1}{2}\,\eta\,L_\tau
\end{align}
In the case of white light, Eq.~\eqref{eq:EpaisseurMaxMontageClassiqueHaidinger} gives, for $\eta=40$, $e_\text{max}=2\;{\rm\mu m}$.

\subsection{Analysis of the interference pattern in white light}
To understand the irradiance distribution that appears on the observation screen using a white light source in Fig.~\ref{fig:MichelsonCondenseurFiltreAntitehrmique}, we first consider each monochromatic wavelength separately. Beforehand, two clarifications are necessary:
\begin{list}{--}{}
\item the thickness $e$ of the equivalent air-blade has been fixed at its maximum value $e_\text{max}=2\;{\rm\mu m}$ to take into account the limitation due to the temporal coherence;
\item the values of the angle of incidence $i$ must be less than or equal to the maximum permitted value $i_\text{max}$ which is equal to $i_\text{max}=0.1$ in the case of the Michelson interferometer model used.
\end{list}

\par
\begin{figure*}[ht]
  \begin{center}
  \hfil
  \subfloat[\label{subfig:DifferentsEclairementsMonochromatiqueMontageClassiqueMichelsonLameDair}]
           {\includegraphics[scale=1]{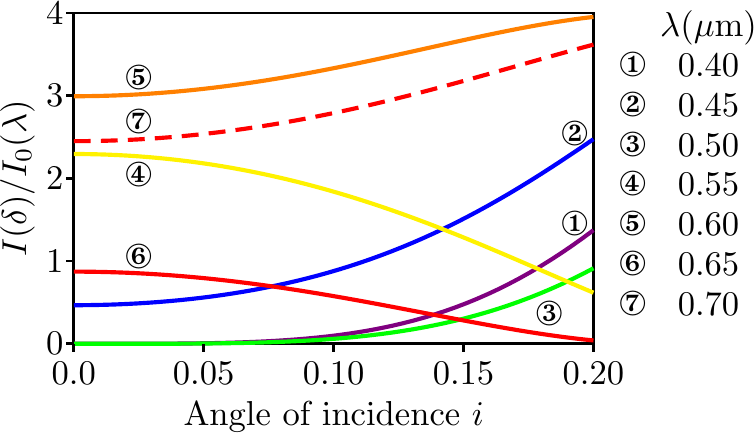}}%
  \hfill
  \subfloat[\label{subfig:DiagrammeLambdaAngleMontageClassiqueMichelsonLameDair}]
           {\includegraphics[width=7cm]{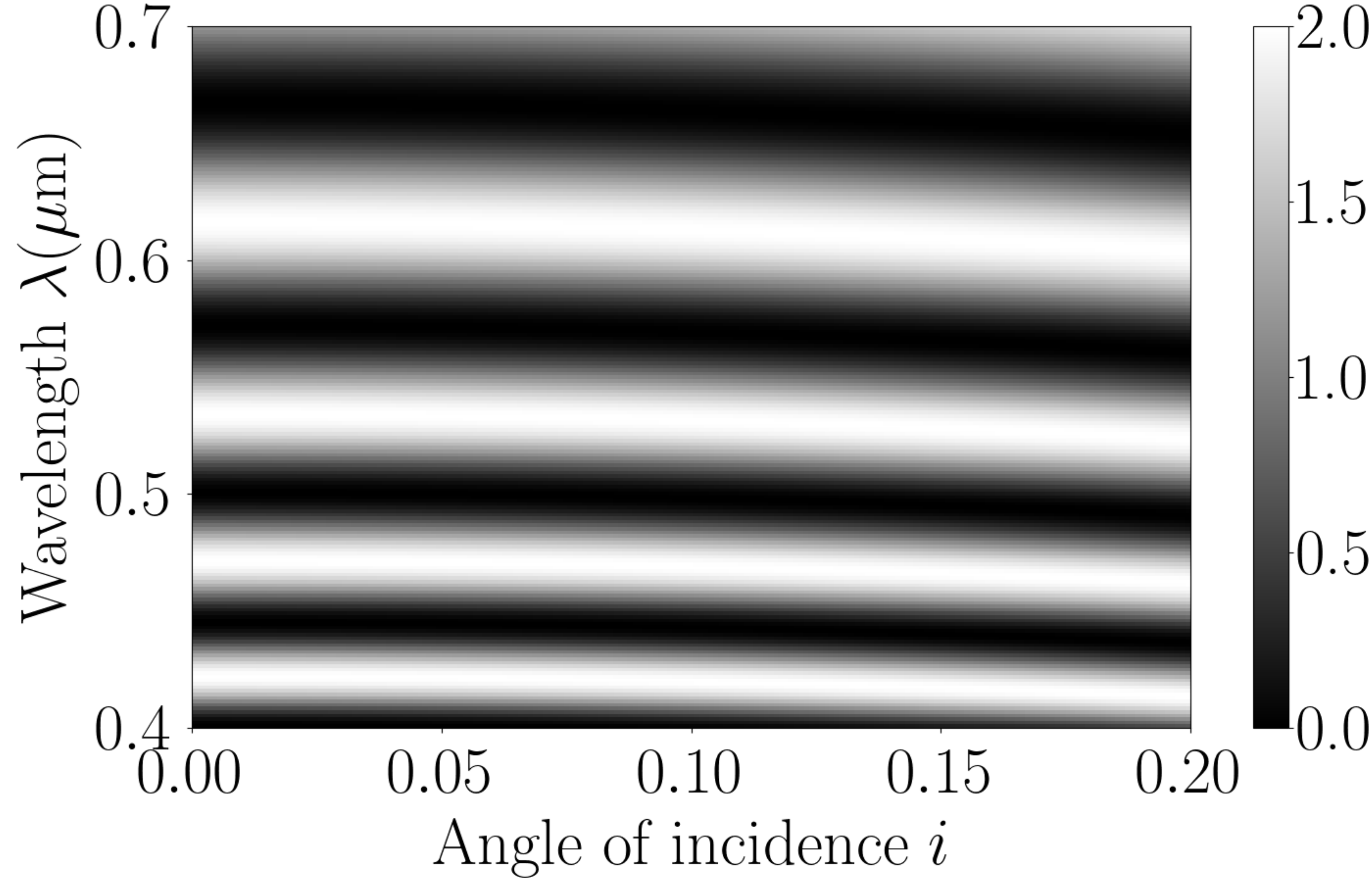}}
  \hfil
  \mbox{}
  \caption{(a) Theoretical curves giving the scaled monochromatic irradiance $I/I_0$ of the screen as a function of the 
           angle of incidence $i$ for different wavelengths $\lambda$ and a thickness $e=2\;{\rm\mu m}$. 
           (b) Theoretical gray-scale map of the scaled monochromatic irradiance versus wavelength and incident angle.
           In practice the field of observation is limited to $i\leqslant i_\text{max}$ with $i_\text{max}=0.1$.
           \label{fig:EclairementAngleLongueurOnde}}
  \end{center}
\end{figure*}

 Figure \ref{subfig:DifferentsEclairementsMonochromatiqueMontageClassiqueMichelsonLameDair} graphs the monochromatic irradiance given by Eq.~\eqref{eq:ExpressionEclairementMonochromatiqueMichelsonClassique} as a function of the position on the screen, indicated by the angle of incidence $i$, for different wavelengths.

\par Figure \ref{subfig:DiagrammeLambdaAngleMontageClassiqueMichelsonLameDair} is a gray-scale map of the irradiance as a function of both incident angle $i$ and wavelength $\lambda$. These two representations clearly demonstrate that, even for the limiting value $e_\text{max}$ of the thickness $e$, the monochromatic irradiances, corresponding to the different wavelengths, do not undergo appreciable variations over the whole permissible field of observation ($0\leqslant i\leqslant i_\text{max}$; $i_\text{max}=0.1$). This lack of variation does not favor obtaining interference rings in the field of observation allowed by the interferometer.

\par Table \ref{tab:RayonsAngulairesPremiersAnneauxHaidingerClassique} gathers the values of the angular radius $i_1$ of the first dark ring for different wavelengths. It shows that, unlike the case of the spectral line, in the case of white light, even filtered using an interference filter of spectral width $\Delta\lambda=10\;{\rm nm}$, the values of $i_1$ are always higher or, at best, close to the limit value of the field of observation $i_\text{max}$. This explains why, with a conventional setup, it is impossible to observe Haidinger rings in white light, whereas one can easily obtain them using spectral line sources.

\begin{table*}[ht]
  \begin{center}
  \begin{ruledtabular}
    \begin{tabular}{r|ccccccccc|c|c}
                             &\multicolumn{9}{c|}{(a)}                    &(b) &(c) \\
    \hline
    $\lambda\mathrm{(\mu m)}$&0.40&0.45&0.50&0.55&0.60&0.65&0.70&0.75&0.80&0.55&0.55 \\
    $i_1(\text{rad})$        &0.45&0.66&0.51&0.60&0.71&0.62&0.78&0.71&0.64&0.08&0.02
    \end{tabular}
    \caption{Angular radius $i_1$ of the first dark ring for: (a) monochromatic light of different wavelengths; (b) 
             white light filtered using an interference filter of bandwidth $\Delta\lambda=10\;{\rm nm}$ centered on the 
             wavelength $\lambda=0.55\;{\rm\mu m}$; and (c) a spectral line of width $1\;{\rm nm}$ centered on the 
             wavelength $\lambda=0.55\;{\rm\mu m}$.
             \label{tab:RayonsAngulairesPremiersAnneauxHaidingerClassique}}
  \end{ruledtabular}
  \end{center}
\end{table*}

\subsection{Comparison with the case of Fizeau fringes\label{ssec:ComparaisonAvecFizeau}}
\begin{figure*}[ht]
  \begin{center}
  \hfil
  \subfloat[\label{subfig:MontageClassiqueMichelsonCoinDair}]
           {\includegraphics[scale=1]{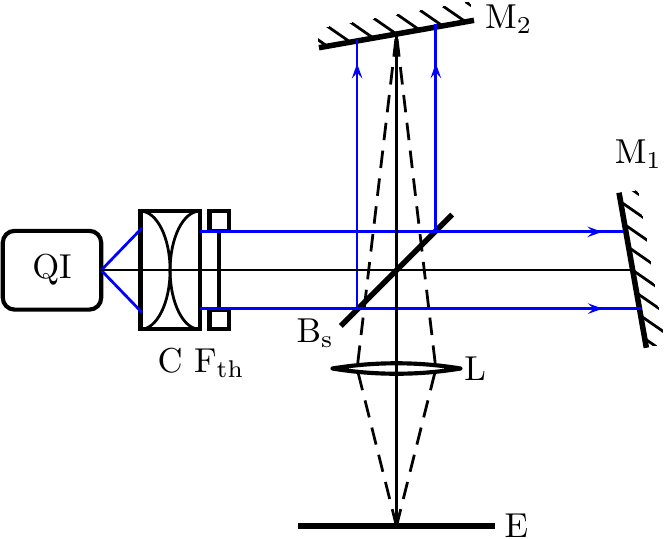}}
  \hfil\hfil
  \subfloat[\label{subfig:CoinDairEquivalenteMichelson}]
           {\includegraphics[scale=1]{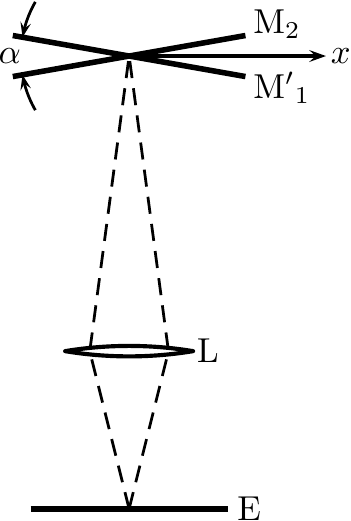}}
  \hfil
  \mbox{}
  \caption{Conventional setup to obtain Fizeau fringes with a Michelson interferometer. Simplified scheme of the 
           Michelson-based device set to an air wedge and illuminated by an almost parallel beam of light (a) and the 
           equivalent air wedge to the Michelson interferometer (b). Dashed lines have been added to show that the screen 
           $\mathrm{E}$ and the air wedge formed by the mirror $\mathrm{M}_2$ and the image $\mathrm{M}’_1$ of the mirror 
           $\mathrm{M}_1$ given by the beam splitter $\mathrm{B_s}$ are conjugated by the lens $\mathrm{L}$. In order to 
           simplify the scheme, the beams reflected by the mirrors $\mathrm{M}_1$ and $\mathrm{M}_2$ are not represented.
           \label{fig:InterferometreDeMichelsonRegleEnCoinDair}}
  \end{center}
\end{figure*}

To test the validity of our criterion for observing interference fringes, we apply it to the case of Fizeau fringes. Unlike Haidinger fringes, Fizeau fringes are easily observable in experiments carried out with the conventional setup based on Michelson interferometer.\citep{Hecht2002,Born1998} The interferometer must be set to an air wedge configuration near the optical contact and illuminated by an almost parallel beam of light. The fringes are rectilinear and localized in the vicinity of the air wedge formed by the mirror $\mathrm{M}_2$ and the image $\mathrm{M}’_1$ of the mirror $\mathrm{M}_1$ given by the beam splitter $\mathrm{B_s}$ (see Fig. \ref{fig:InterferometreDeMichelsonRegleEnCoinDair}).

\par The observation of the fringes is still dependent on two constraints:
\begin{list}{--}{}
\item the field of observation of the interferometer which, in this case, is limited by the diameter $\phi$ of the mirrors of the interferometer, $|x|\leqslant\phi/2$, with $\phi=4\;{\rm cm}$ for the model used;
\item the ratio (see Eq.~\eqref{eq:ExpressionGalContrainteDdmCoherenceTemporelle}) of the temporal coherence length to the optical path difference $\delta=2\,\alpha\,x$, where $\alpha$ is the angle of the air wedge and $x$ is the position in the interference field shown in Fig.~\ref{subfig:CoinDairEquivalenteMichelson}, counted from the edge of the air wedge:
      \begin{align}
      |x|\leqslant\frac{\eta\,L_\tau}{2\,\alpha}.
      \end{align}
\end{list}

\par It can already be noted that, unlike in the case of the Haidinger rings, the two constraints contribute in the same way. Indeed, whereas in the case of Haidinger fringes, the weak temporal coherence imposes to work in the vicinity of the optical contact ($\ell_1\approx\ell_2$) and widens the interference rings; while the size and arrangement of the optical elements constituting the interference device limit the field of observation to low angles; in the case of the Fizeau fringes the two constraints impose a limitation of the field of observation. Thus, according to the value of the angle $\alpha$, the limitation is imposed either by the temporal coherence (case of the large values of $\alpha$) or by the size of the mirrors (case of the small values of $\alpha$).

\par The rectilinear fringes obtained are equidistant from eath other. The separation of adjacent bright or dark fringes is given by the classical relation:
\begin{align}
\label{eq:InterfrangeFizeau}
i=\frac{\lambda}{2\,\alpha}.
\end{align}
The number of monochromatic fringes that can be observed is then given by:
\begin{align}
N=\frac{1}{\lambda}\,\mathrm{Min}\left({\eta\,L_\tau,\alpha\,\phi}\right)
\end{align}
$\mathrm{Min}(x,y)$ denotes the smaller value of $x$ or $y$. In practice, the number of fringes observable for a given wavelength is limited by  temporal coherence. Indeed, the limitation by the size of the mirrors becomes predominant for angles $\alpha$ lower than a limit angle:
\begin{align}
\alpha_\text{lim}=\frac{\eta\,L_\tau}{\phi}.
\end{align}
Numerically, in the case of white light with $L_\tau=0.1\;{\rm\mu m}$, $\eta=40$ and for mirrors of diameter $\phi=4\;{\rm cm}$, we obtain $\alpha_\text{lim}=10^{-4}$.

\par Table \ref{tab:NombreFrangesMonochromatiquesObservablesMichelsonCoinDair} gives the value of the number of fringes observable in monochromatic light for different wavelengths of the visible spectrum. As a result, the irradiance undergoes a sufficient number of oscillations over the extent of the observation field so that a considerable number of fringes of different colors can be obtained by additive mixing (additive color).

\begin{table*}[ht]
  \begin{center}
  \begin{ruledtabular}
    \begin{tabular}{r|ccccccccc}
    $\lambda(\mathrm{\mu m})$&0.40&0.45&0.50&0.55&0.60&0.65&0.70&0.75&0.80\\
    \hline
    $N$                      &10.0&8.9 &8.0 &7.3 &6.7 &6.2 &5.7 &5.3 &5.0
    \end{tabular}
    \caption{Number of Fizeau fringes observable using a conventional Michelson interferometer setup, in monochromatic 
             light at different wavelengths $\lambda$ constituting white light.
             \label{tab:NombreFrangesMonochromatiquesObservablesMichelsonCoinDair}}
  \end{ruledtabular}
  \end{center}
\end{table*}

\section{Modified setup\label{sec:HaidingerModifie}}
By using the criterion for observing interference fringes (Eq.~\eqref{eq:ExpressionGalContrainteDdmCoherenceTemporelle}), we learned why it is so difficult to obtain Haidinger rings in white light using the conventional setup based on a Michelson interferometer. Before we show that this same criterion provides for the easy observation of Haidinger rings with the modified setup, we present a brief description of such a setup.

\subsection{Arrangement of the optical elements and theoretical relations\label{ssec:DispositionElementsRelationsTheoriques}}
\begin{figure*}[ht]
  \begin{center}
  \hfil
  \subfloat[\label{subfig:MontageModifieMichelsonLameDair}]
           {\includegraphics[scale=1]{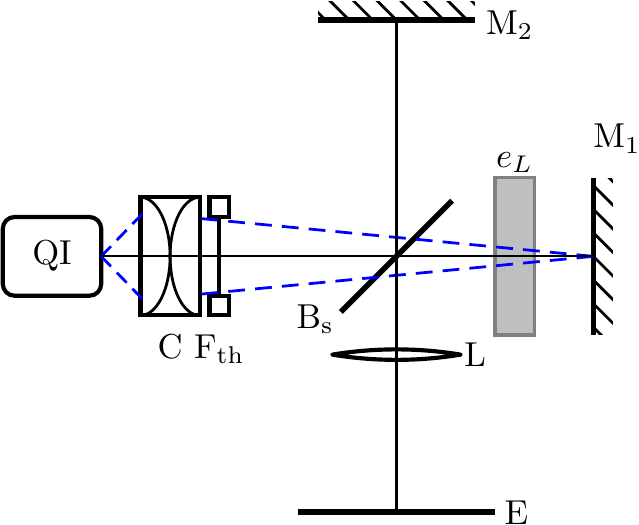}}
  \hfil
  \subfloat[\label{subfig:SchemaEquivalentMontageModifieMichelson}]
           {\includegraphics[scale=1]{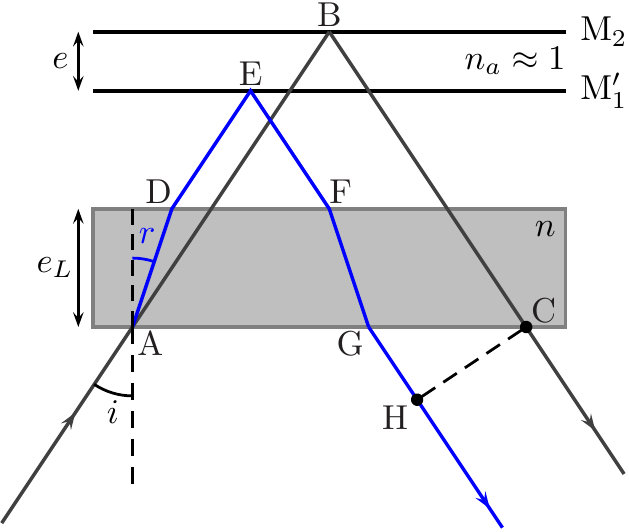}}
  \hfil
  \mbox{}
  \caption{Modified interference setup used to obtain Haidinger-Michelson rings in white light (a) and its equivalent 
           scheme with double parallel-face blade (b). $\text{QI}$ is a quartz-iodine source; $\text{C}$ is a condenser or 
           a converging lens for illuminating the device with a convergent light beam; $\mathrm{F_{th}}$ is an optional 
           thermal filter; $\mathrm{B_s}$ represents the beam splitter; $\mathrm{{M_1}}$ is the moving mirror; and 
           $\mathrm{M_2}$ the fixed mirror. $\mathrm{L}$ is a converging lens allowing the image of the 
           interference pattern to appear on the observation screen $\mathrm{E}$ placed in its focal plane.
           \label{fig:MontageMichelsonModifie}}
  \end{center}
\end{figure*}

The modified setup is shown in Fig.~\ref{fig:MontageMichelsonModifie} and uses a conventional Michelson interferometer set as close as possible to the optical contact in the configuration called “parallel-faces air blade." A transparent glass slide with parallel faces is introduced in front of the mirror $\mathrm{M}_1$, as shown in Fig. \ref{subfig:MontageModifieMichelsonLameDair}.

\par Let $e_L$ and $n$ denote respectively the thickness and the refractive index of the glass slide. For the calculation of the optical path difference, the equivalent diagram shown in Fig. \ref{subfig:SchemaEquivalentMontageModifieMichelson} can be used. Everything happens as if there is interference between the wave reflected by the mirror $\mathrm{M}_2$ and that transmitted by the glass slide before being reflected by the mirror $\mathrm{M}’_1$ and then transmitted a second time in the opposite direction by the glass slide.

\par The optical path difference is then given by:
\begin{align}
\delta=(ABC)-(ADEFGH),
\end{align}
which becomes, after some calculations:
\begin{align}
\label{eq:ExpressionDdmMontageModifieSansApproximation}
\delta=2\,(e+e_L)\cos(i)-2\,n\,e_L\cos(r),
\end{align}
where $r$ is the internal refractive angle within the glass slide, obtained from the angle of incidence $i$ using the second Descartes-Snell law for refraction $\sin(i)=n\sin(r)$.

\par In the case of the modified setup, monochromatic illumination produces irradiance a the observation screen given by Eq.~\eqref{eq:ExpressionEclairementMonochromatiqueMichelsonClassique} while white light illumination produces irradiance  given by Eq.~\eqref{eq:ExpressionEclairementLumiereBlancheMichelsonClassique}. In these expressions, it is necessary to use Eq.~\eqref{eq:ExpressionDdmMontageModifieSansApproximation} for the optical path difference $\delta$.

\par In the context of the paraxial approximation, the angles $i$ and $r$ are small enough that Eq.~\eqref{eq:ExpressionDdmMontageModifieSansApproximation} is approximated by:
\begin{align}
\delta=2\,\left[{e-(n-1)\,e_L}\right]-\left({\frac{n-1}{n}\,e_L+e}\right)\,i^2.
\end{align}
This expression clearly shows the additional optical path difference $-2\,(n-1)\,e_L$ introduced in the center of the interference pattern ($i = 0$) by the glass slide. In general, this additional optical path difference is sufficient to move the setup away from the optical contact. To recover the interference conditions, it is necessary to compensate for this additional optical path difference by means of an adequate translation $e$ of the moving mirror. However, taking into account the dispersion of the glass constituting the slide, this compensation can be achieved only for a single wavelength $\Lambda_0$ at a time, such that:
\begin{align}
\label{eq:RelationCompensationDDM}
e=[n(\Lambda_0)-1]\,e_L.
\end{align}
The optical path difference then becomes for any wavelength $\lambda$:
\begin{align}
\delta=2\,(n_0-n)\,e_L-e_L\,\frac{n\,n_0-1}{n}\,i^2,
\end{align}
where $n_0=n(\Lambda_0)$ represents the value of the refractive index of the parallel-faces glass slide for the compensation wavelength $\Lambda_0$. Under these conditions, the angular radius $i_m(\lambda)$ of the dark ring ($m$) from the center becomes:
\begin{align}
i_m(\lambda)=\sqrt{\frac{n\,\lambda}{(n\,n_0-1)\,e_L}\,(m+\epsilon)},
\end{align}
where $\epsilon=p_0-\mathrm{E}(p_0)$. $\mathrm{E}(p_0)$ is the integer part of the interference order $p_0$ in the center of the interference pattern ($i=0$), given by:
\begin{align}
p_0=2\,(n_0-n)\,\frac{e_L}{\lambda}.
\end{align}

\par
\begin{figure*}[ht]
  \begin{center}
  \hfil
  \subfloat[\label{subfig:DifferentsEclairementsMonochromatiqueMontageModifieMichelsonLameDair}]
           {\includegraphics[scale=1]{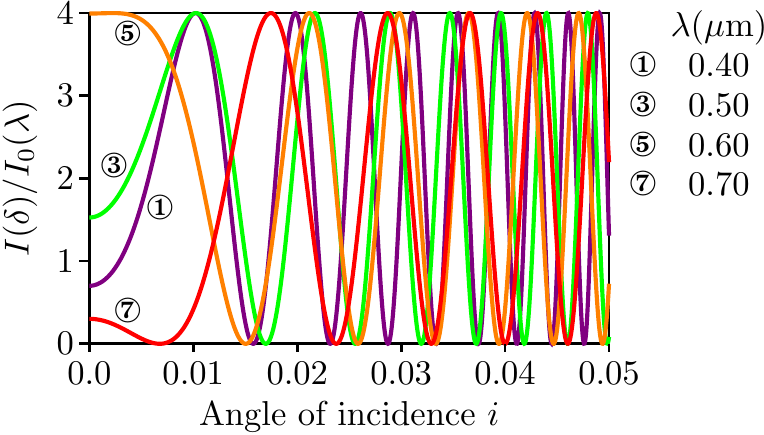}}
  \hfill
  \subfloat[\label{subfig:VaraiationEclairementEnFonctionLongueurOndeAuCentreMontageModifieMichelsonLameDairDensity}]
           {\includegraphics[width=7cm]{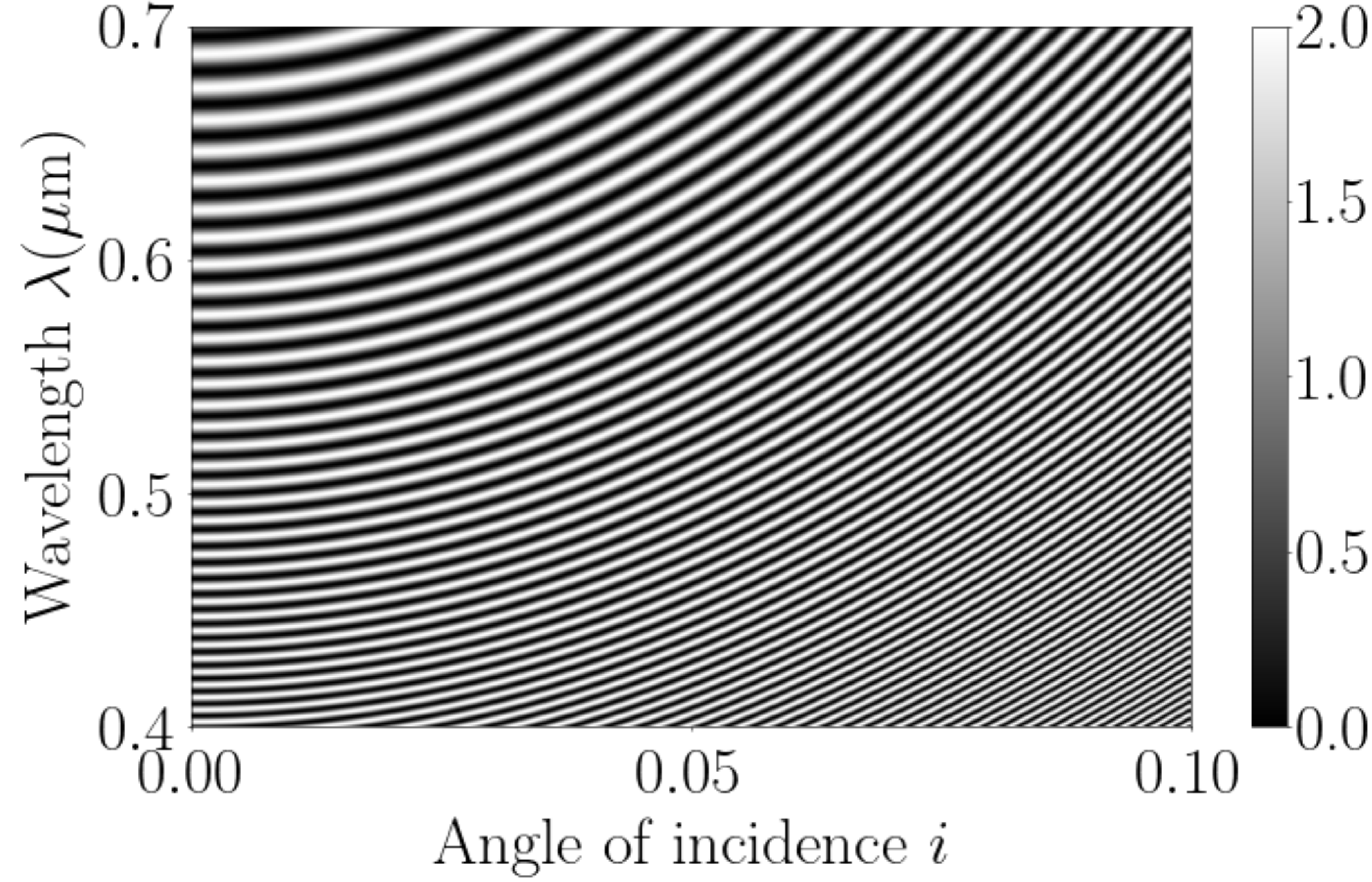}}
  \hfil
  \mbox{}
  \caption{Monochromatic irradiance at the screen in the case of the modified setup: (a) irradiance 
           versus incident angle for several wavelengths $\lambda$ and (b) a grays-cale irradiance map versus wavelength 
           and incident angle. The thickness of the glass slide is $e_L=1.5\;{\rm mm}$, the compensation wavelength is 
           $\Lambda_0=1\;{\rm\mu m}$ corresponding to a translation of the moving mirror of $e=814\;{\rm\mu m}$ from 
           the position corresponding to the optical contact. For reasons of visibility of the graphical representation 
           (a), we have only represented $4$ monochromatic irradiance curves and the range of incidence angles $i$ has 
           been limited to $0.00\;-\;0.05$.
           \label{fig:EclairementsMonochromatiquesMichelsonModifieLameDair}}
  \end{center}
\end{figure*}

 Figure \ref{subfig:DifferentsEclairementsMonochromatiqueMontageModifieMichelsonLameDair} provides a graph of the monochromatic illumination of the screen as a function of the angle of incidence $i$ for some wavelengths $\lambda$ and Fig.~\ref{subfig:VaraiationEclairementEnFonctionLongueurOndeAuCentreMontageModifieMichelsonLameDairDensity} shows an angle-wavelength diagram giving the light intensity in gray density level for a glass slide of thickness $e_L=1.5\;{\rm mm}$, and a compensation wavelength $\Lambda_0=1\;{\rm\mu m}$ corresponding to a translation of the moving mirror of $e=814\;{\rm\mu m}$ from the position corresponding to the optical contact ($\ell_1=\ell_2$). Unlike the case of the conventional setup (see Fig.~\ref{fig:EclairementAngleLongueurOnde}), the irradiance at a particular wavelength oscillates several times in the field of observation ($i\leqslant i_\text{max}$) with a proper period for each monochromatic radiation. The result is that, for each value of the angle of incidence $i$, there is a superposition of different wavelengths and different intensities. By additive mixing, a trichromatic detector, like the human eye, then perceives circular isochromatic fringes ($i=\text{constant}$) in the interference field.

\subsection{Experimental procedure\label{ssec:ExperimentalProtocol}}
To obtain the Haidinger-Michelson rings in white light (see Fig. \ref{subfig:MontageModifieMichelsonLameDair}), one must follow a precise experimental procedure.
\begin{enumerate}
\item In the absence of the parallel-faces glass slide, the Michelson interferometer is set in the air wedge configuration as close as possible to the optical contact and illuminated with a quasi-parallel beam of white light to obtain rectilinear fringes localized near the wedge.
\item The rectilinear fringes are projected onto the observation screen $\mathrm{E}$ using the lens $\mathrm{L}$ and the angle $\alpha$ of the wedge is progressively decreased so as to increase the distance between adjacent fringes (Eq.~\eqref{eq:InterfrangeFizeau}) until colored tints, thus obtained, occupy almost the entire interference field on the observation screen.
\item The parallel-faces glass slide is then introduced in front of the moving mirror $\mathrm{M}_1$. This is generally accompanied by a disappearance of the colored tints because of the additional optical path difference introduced by the glass slide.
\item The movable mirror is translated toward the glass slide to compensate for the additional optical path difference introduced by the slide and thus to make the colored fringes reappear.
\item Finally, the condenser $\mathrm{C}$ is moved toward the beam splitter in order to illuminate the interferometer with a convergent beam, and the observation screen $\mathrm{E}$ is placed in the focal plane of the lens $\mathrm{L}$.
\end{enumerate}

\begin{figure*}[ht]
  \begin{center}
    \hfil
    \subfloat[$e=865\;{\rm\mu m}$]
             {\includegraphics[width=4.5cm,height=3cm]{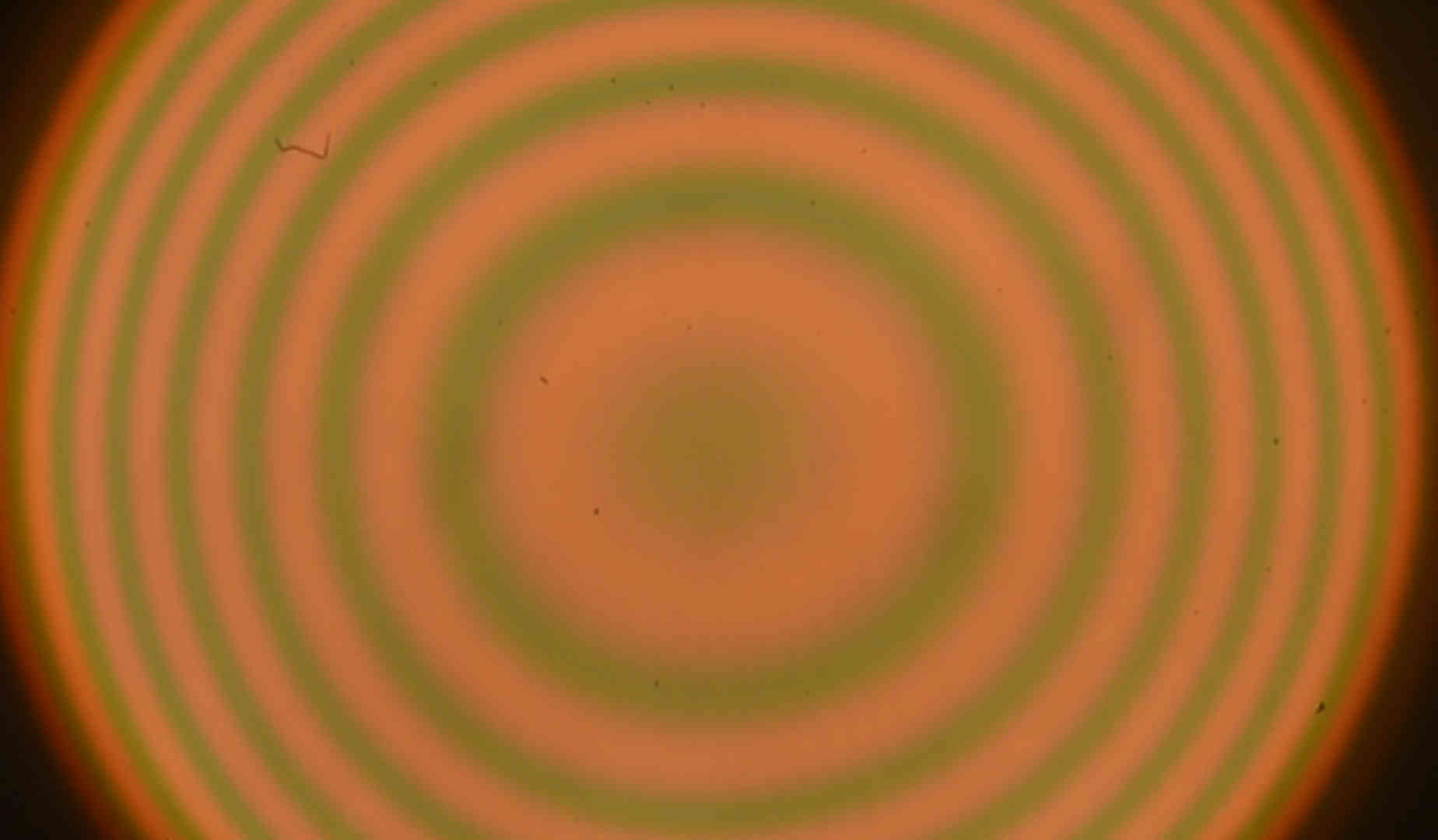}}
    \hfil
    \subfloat[$e=854\;{\rm\mu m}$]
             {\includegraphics[width=4.5cm,height=3cm]{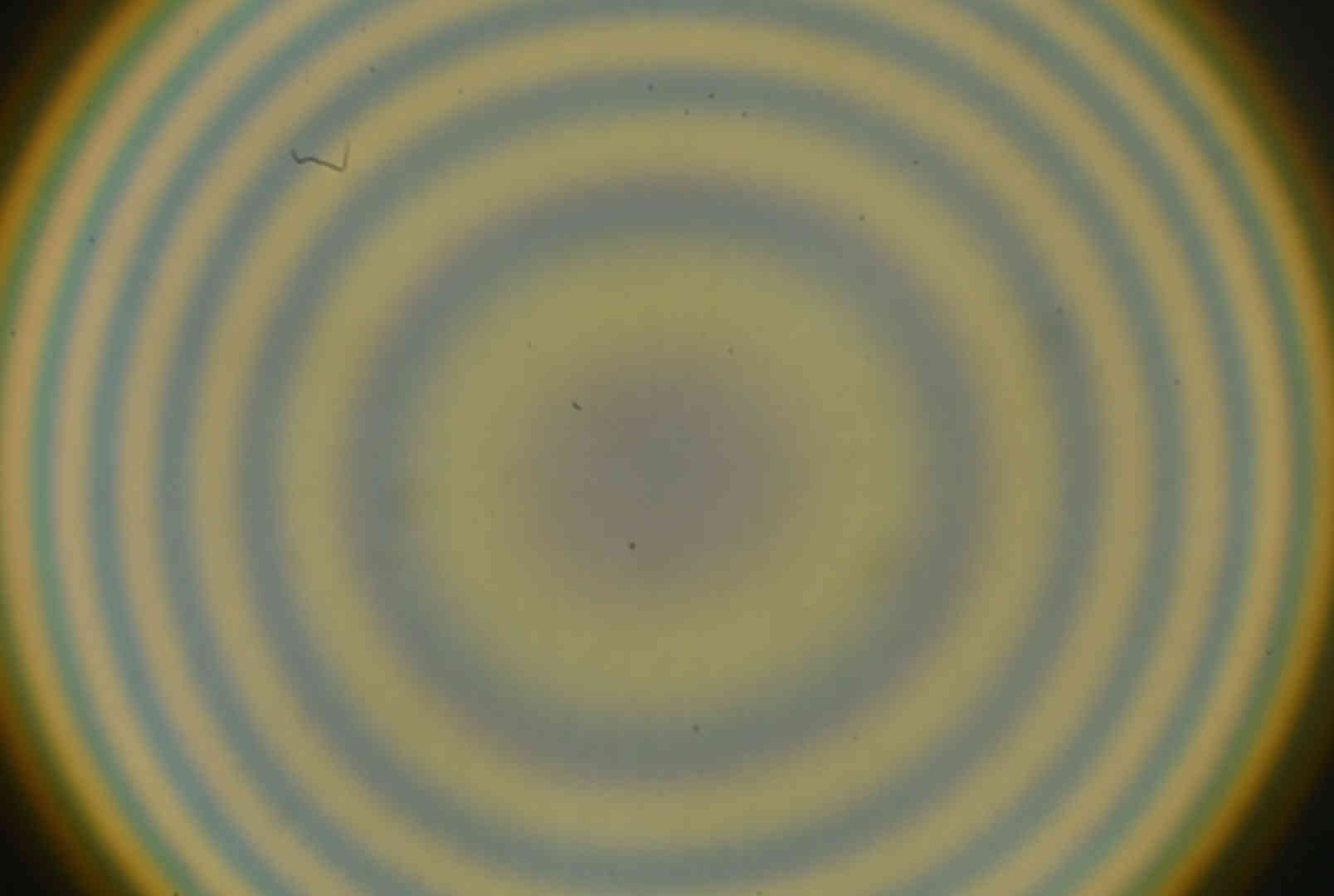}}
    \hfil
    \subfloat[$e=849\;{\rm\mu m}$]
             {\includegraphics[width=4.5cm,height=3cm]{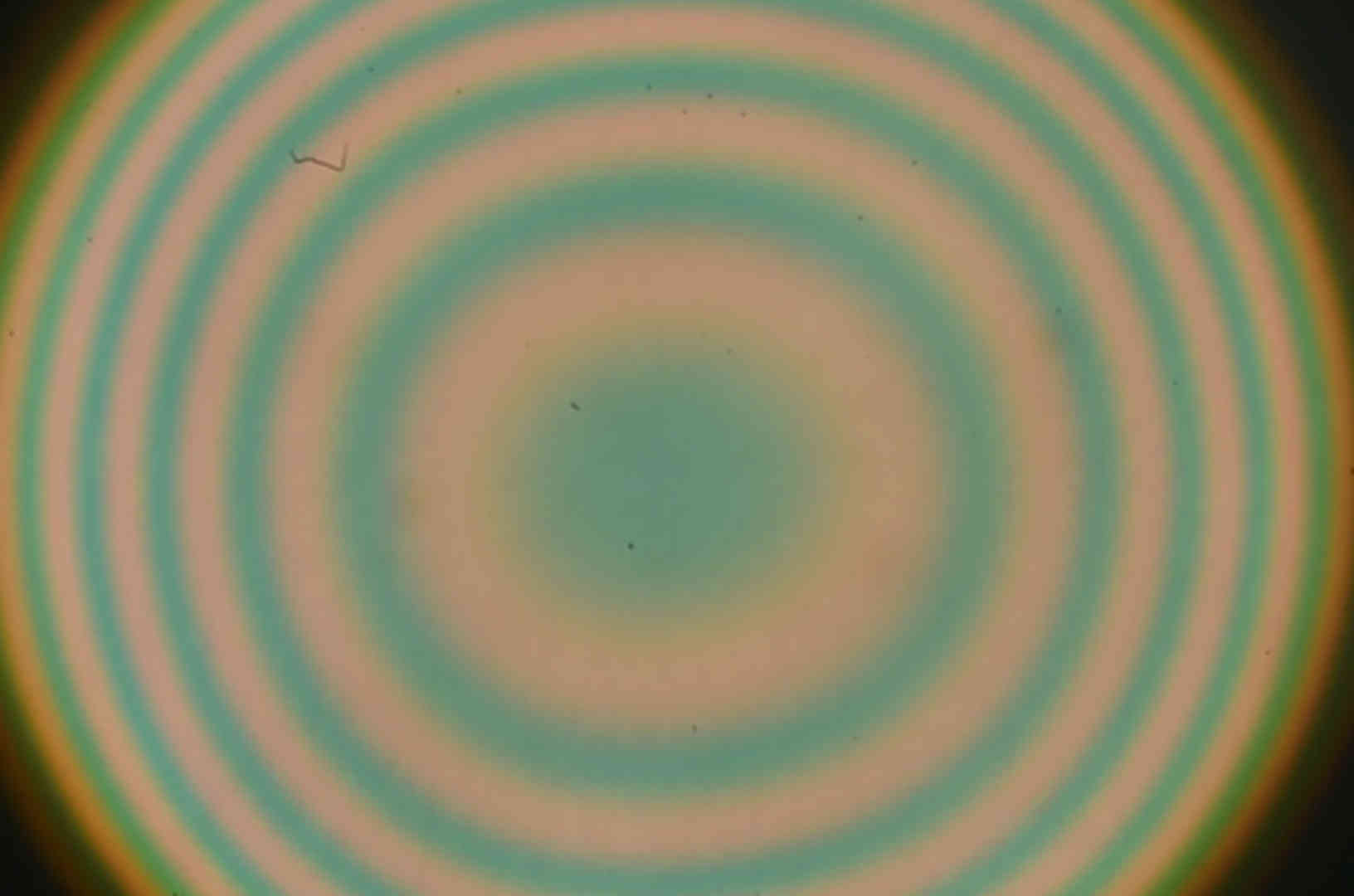}}
    \hfil
    \mbox{}
    \caption{Haidinger-Michelson rings obtained using the modified setup, for three different displacements $e$ of the 
             moving mirror from the position corresponding to the optical contact (initially obtained in the absence of 
             the faces glass slide).
             \label{fig:AnneauHaidingerMontageMichelsonModifie}}
  \end{center}
\end{figure*}

\subsection{Experimental results}

\par Figure \ref{fig:AnneauHaidingerMontageMichelsonModifie} gives some examples of Haidinger-Michelson rings in white light experimentally obtained with the modified Michelson interferometer setup, for three different displacements of the moving mirror from the position corresponding to the optical contact (initially obtained in the absence of the glass slide).

\par
\begin{figure*}[ht]
  \begin{center}
    \hfil
    \subfloat[Red filter]
             {\includegraphics[height=4cm]{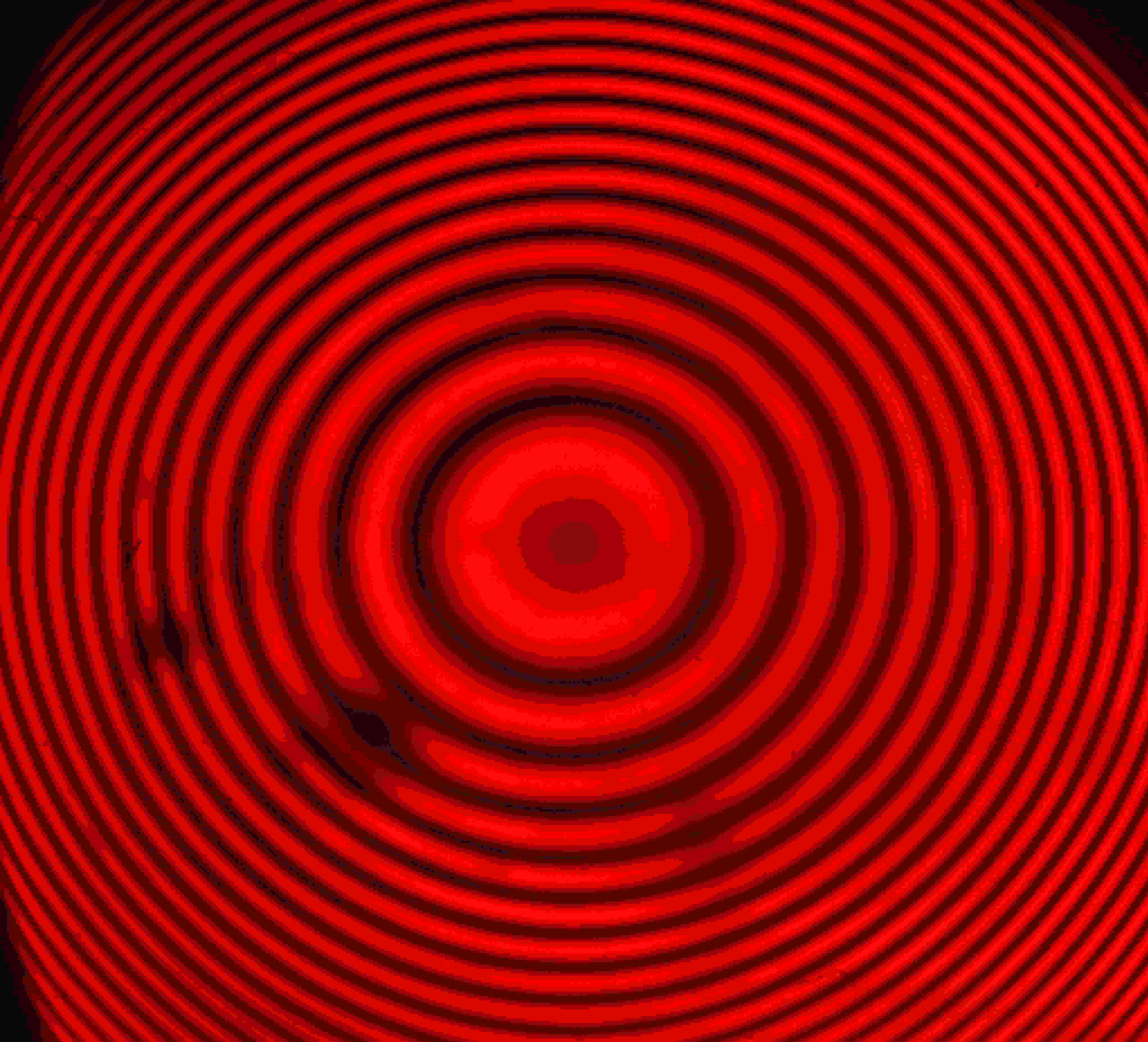}}
    \hfil
    \subfloat[Yellow filter]
             {\includegraphics[height=4cm]{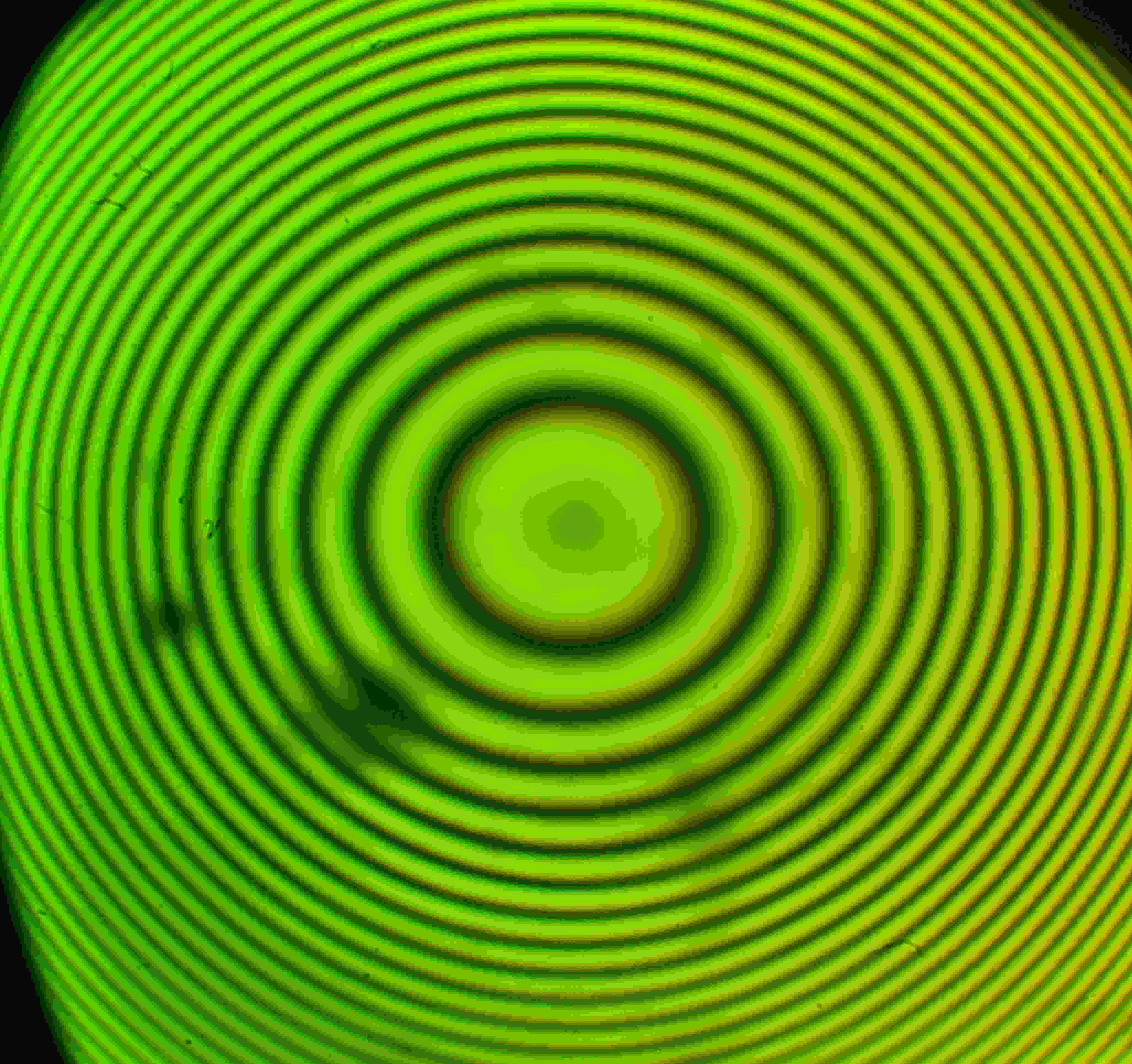}}
    \hfil
    \subfloat[Green filter]
             {\includegraphics[height=4cm]{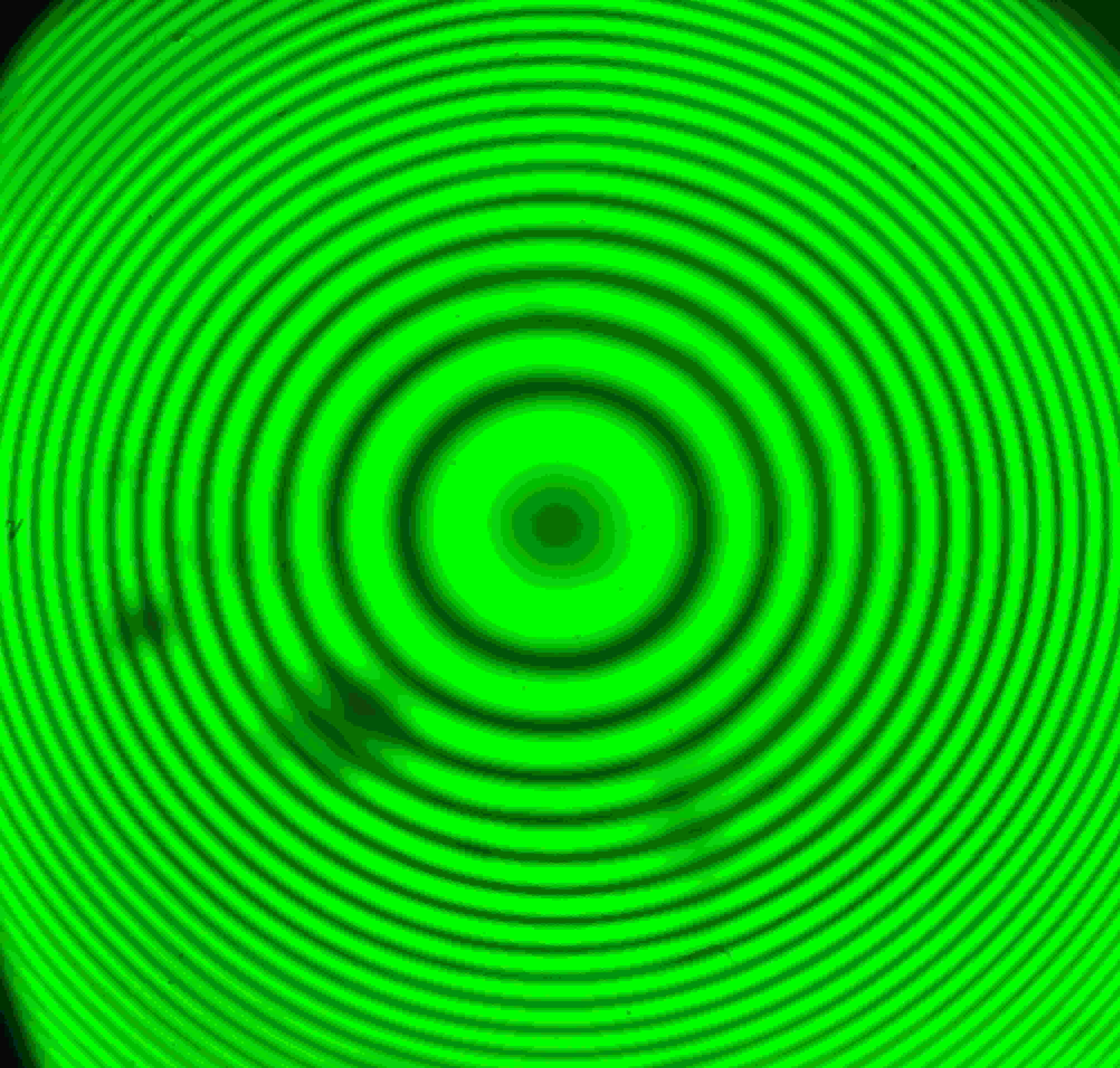}}
    \hfil
    \mbox{}
    \caption{Haidinger-Michelson rings obtained with different interference filters of the same bandwidth 
             $\Delta\lambda=10\;{\rm nm}$:
             (a) red filter ($\lambda=640\;{\rm nm}$), 
             (b) yellow filter ($\lambda=578\;{\rm nm}$) and 
             (c) green filter ($\lambda=546\;{\rm nm}$).
             \label{fig:AnneauxHaidingerMontageModifieMichelsonLameDairFiltreInterferentiel}}
  \end{center}
\end{figure*}

 The modified setup (see Fig. \ref{subfig:MontageModifieMichelsonLameDair}) also makes it possible to obtain the Haidinger-Michelson rings with interference filters. Figure \ref{fig:AnneauxHaidingerMontageModifieMichelsonLameDairFiltreInterferentiel} gives some examples of interference patterns obtained with three interference filters of bandwidth $\Delta\lambda=10\;{\rm nm}$, respectively red ($\lambda=640\;{\rm nm}$), yellow ($\lambda=578\;{\rm nm}$) and green ($\lambda=546\;{\rm nm}$).

\section{Spectral interferometry\label{sec:InterferometrieSpectrale}}
In this part, we use the modified setup to perform spectral interferometry by analyzing the interferometric signal corresponding to the higher-order white light obtained at the output of the interferometer for a sufficiently large optical path difference. In our case, the analysis is carried out using a diffraction grating spectrometer with a typical resolution of 1 nm. The sensor consists of an optical fiber whose input is placed at the center ($i=0$) of the interference pattern. Equation \eqref{eq:ExpressionEclairementMonochromatiqueMichelsonClassique} shows that the intensity $I(\delta_0)$ at the center of interference pattern vanishes for wavelengths $\lambda_k$ satisfying $\delta_0=k\,\lambda_k$, where $k\in\mathbb{Z}$ and $\delta_0$ the optical path difference at the center ($i=0$) of interference pattern. The wavelengths thus extinguished correspond to dark spectral fringes, or splines, in the spectrum of the analyzed light.

\par For a translation $e$ of the moving mirror from the position of optical contact, corresponding to the compensation wavelength $\Lambda_0$ (see sec.~\ref{ssec:DispositionElementsRelationsTheoriques}), the optical path difference $\delta_0$ in the center of the interference pattern, for the wavelength $\lambda$, can be written:
\begin{align}
\delta_0=\pm2\,(n_0-n)\,e_L.
\end{align}
The sign $\pm$ allows us to define a positive optical path difference $\delta_0$ and thereafter to distinguish two cases. The first one, with the “$-$" sign, corresponds to dark spectral fringes (or splines) located before the compensation wavelength; that is $\lambda_k<\Lambda_0$ and $n>n_0$. The second case, with the “$+$" sign, corresponds to splines located after the compensation wavelength; that is $\lambda_k>\Lambda_0$ and $n<n_0$.

\par If the wavelength difference between successive splines is sufficiently small that we can neglect the variations of index between such fringes, then we find:
\begin{align}
\label{eq:IndiceRefractionCanneluresDeuxCas}
n(\lambda_k)=\left\{
                    \begin{array}{ll}
                    \displaystyle1+\frac{e}{e_L}+\frac{\lambda_k\,\lambda_{k+1}}{2\,e_L\,|\lambda_k-\lambda_{k+1}|}&\text{for}\;\lambda_k<\Lambda_0\\[5mm]
                    \displaystyle1+\frac{e}{e_L}-\frac{\lambda_k\,\lambda_{k+1}}{2\,e_L\,|\lambda_k-\lambda_{k+1}|}&\text{for}\;\lambda_k>\Lambda_0.
                    \end{array}
             \right.
\end{align}

\par
\begin{figure*}[t]
  \begin{center}
    \hfil
    \subfloat[\label{subfig:SpectreCannele}]
             {\includegraphics[width=7cm,height=7cm]{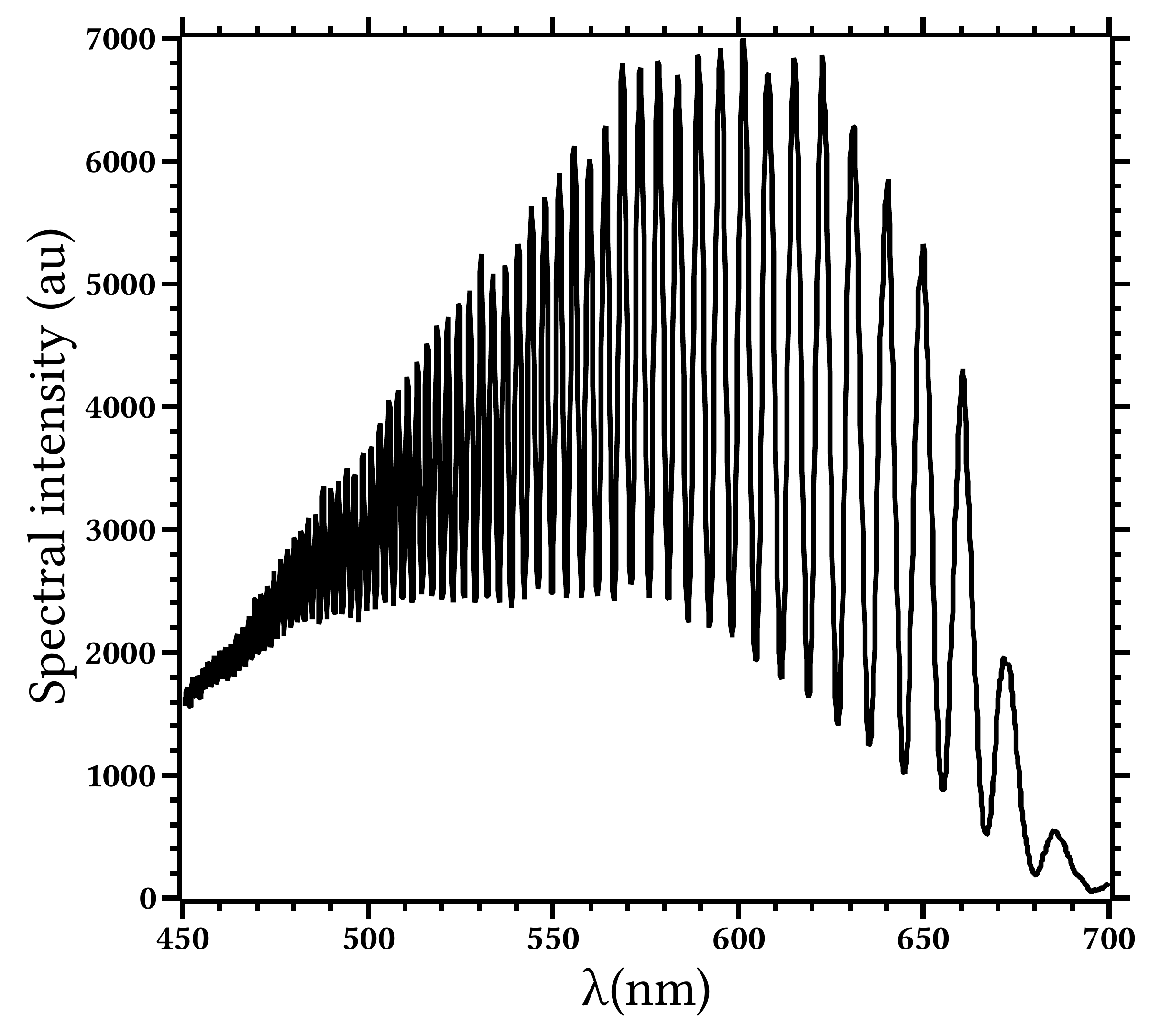}}
    \hfil
    \subfloat[\label{subfig:Sellmeier}]
             {\includegraphics[width=7cm,height=7cm]{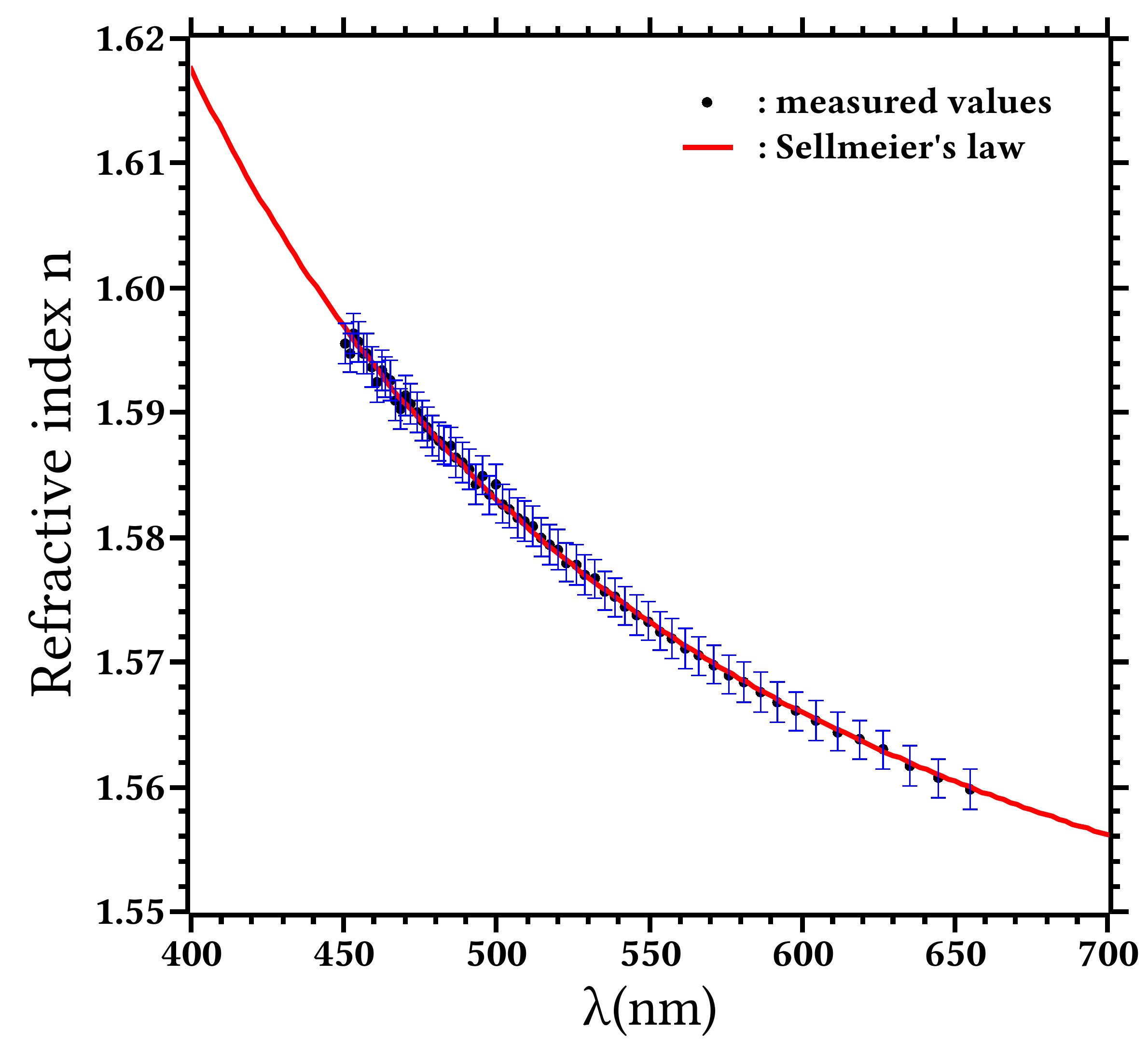}}
    \hfil
    \mbox{}
    \caption{Determination of the refractive index of the parallel-faces glass slide from the interference spectrum (a) 
             obtained with a glass slide of a thickness $e_L=1.5\;{\rm mm}$ for a translation of the moving mirror 
             $e=821\;{\rm\mu m}$, and the fit of refractive index measurements of the glass slide using Sellmeier's law 
             (b).\label{fig:SpectreCanneleFitSellmeier}}
  \end{center}
\end{figure*}

 Figure \ref{subfig:SpectreCannele} gives the spectrum obtained with a glass slide with a thickness $e_L=1.5\;{\rm mm}$ for a translation $e=821\;{\rm\mu m}$ of the movable mirror of the Michelson interferometer from the position corresponding to the optical contact in the absence of the glass slide. The translation was carried out so as to locate the compensation wavelength $\Lambda_0$ in the infrared region. Equation \eqref{eq:IndiceRefractionCanneluresDeuxCas}, corresponding to the case $\lambda_k<\Lambda_0$, then makes it possible to deduce the values of the refractive index $n$ for the different wavelengths corresponding to the splines. Figure \ref{subfig:Sellmeier} gives the graphical representation of the refractive index thus determined as a function of the wavelength $\lambda$. This figure also gives the fit of the values thus determined by Sellmeier's law~\citep{Born1998,Sellmeier1872} with a single absorption resonance, of a force $A$ and a characteristic wavelength $\lambda_0$:
\begin{align}
n^2=1+\frac{A \lambda^2}{\lambda^2-\lambda_0^2}.
\end{align}

The fitting procedure\citep{Peck1983} was performed using the Levenberg-Marquardt algorithm.\citep{QtiPlot,Levenberg1944,Marquardt1963} Figure \ref{subfig:Sellmeier} shows the very good agreement between the experimental results and the Sellmeier model. Table \ref{tab:ResultatsFitSellmeier} gives the values of the Sellmeier model parameters deduced from its fit to the experimental measurements. These results show that the absorption resonance responsible for the dispersion is located in the near ultraviolet range. From this one can deduce the extrapolated value in the infrared $n_\text{IR}$ of the refractive index of the glass slide ($n_\text{IR}=\displaystyle\lim_{\lambda\rightarrow+\infty}n(\lambda)=\sqrt{1+A}$). This allows also to obtain the values of the refractive index for the whole wavelengths in the UV-visible-Near infrared spectrum; especially for those tabulated in handbooks.\citep{CRC2009} The values thus determined are very closed to those of BaK glass type, or Barium Crown glass.

\begin{table*}[ht]
  \begin{center}
  \begin{ruledtabular}
    \begin{tabular}{cccc}
    $R^2$  &$\lambda_0(\mathrm{nm})$&$A$             &$n_\text{IR}$\\
    \hline
    $0.998$&$164.5\pm0.4$          &$1.3431\pm0.0008$&$1.5307\pm0.0003$
    \end{tabular}
    \caption{Fitting results of the refractive index measurements by the Sellmeier model with a single absorption 
             resonance of force $A$ and a characteristic wavelength $\lambda_0$.\label{tab:ResultatsFitSellmeier}}
    \end{ruledtabular}
    \end{center}
\end{table*}

\section{Conclusion}
Our results explain why it is impossible to observe Haidinger rings in white light with the usual Michelson interferometer setup. The weak temporal coherence of white light widens the interference rings, while the size and separation of the optical elements constituting the interference device limit the field of observation to a point where the rings cannot be observed.

\par The analysis of the white-light interferometer prepared us to understand how a modified setup of the Michelson interferometer overcomes the restrictions imposed by the low coherence of white light. The modified setup has a glass slide in one of the two arms of the interferometer and the additional path difference that it generates is compensated by a translation of the moving mirror of the interferometer.

\par For sufficiently large optical path differences, the modified setup makes it possible to determine the wavelength dependence of the refractive index of the glass slide over the whole visible spectrum. This determination is performed by measuring the wavelengths corresponding to the spectral dark fringes (splines). The fitting of the obtained results  to Selmeier's law gives an estimate of the resonant wavelength characteristic of the oscillator model used as well as the extrapolated value of the refractive index in the infrared. We demonstrated this technique and found a characteristic wavelength $\lambda_0=(164.5\pm0.4)\,\mathrm{nm}$ with a force $1.3431\pm0.0008$ and an IR index $n_\text{IR}=1.5307\pm0.0003$.

\begin{acknowledgments}
The authors would like to thank Mohamed Chafi, Rodolphe Heyd and Jean-Pierre Lecardonnel for their comments and suggestions on a preliminary version of this manuscript. We would also like to thank Noureddine Bendouqi and Abdelkader Outzourhit who proofread the article and really contributed to its enrichment and improvement by their remarks and suggestions. Our thanks also go to the referees who, through their remarks and suggestions, have contributed to considerably improving the quality of this article.
\end{acknowledgments}

\end{document}